\documentstyle[12pt, epsfig]{article}     
     
\begin{document}     
     
\vspace*{-2cm} \renewcommand{\thefootnote}{\fnsymbol{footnote}}     
\begin{flushright}     
hep-ph/0004056\\     
PSI PR-00-07\\ 
April 2000\\     
\end{flushright}     
\vskip 65pt     
     
\begin{center}     
{\Large {\bf Subleading Sudakov logarithms in electroweak high energy processes to
all orders
}}\\[0pt]     
\vspace{1.2cm}    
{Michael Melles\footnote{{\bf     
Michael.Melles@psi.ch}} }\\     
\vspace{10pt}   
  
{Paul Scherrer Institute (PSI), CH-5232 Villigen, Switzerland. }  
\end{center}

\vspace{60pt}     
\begin{abstract}     
In future collider experiments at the TeV scale, large logarithmic corrections originating
from massive boson exchange can lead to significant corrections to observable cross sections.
Recently double logarithms of the Sudakov-type were resummed for spontaneously
broken gauge theories and
found to exponentiate. In this paper we use the virtual contributions to
the Altarelli-Parisi splitting functions to obtain the next to leading order kernel of
the infrared evolution equation in
the fixed angle scattering regime at high energies where particle masses can be neglected. 
In this regime the virtual corrections can be described by a generalized renormalization
group equation with infrared singular anomalous dimensions.
The results are valid for virtual
electroweak corrections to fermions and
transversely polarized vector bosons with an arbitrary number of external lines. 
The subleading terms are found to exponentiate as well and are related to external lines, allowing
for a probabilistic interpretation in the massless limit.
For $Z$-boson and $\gamma$ final states our approach leads 
to exponentiation with respect to each amplitude
containing the fields of the unbroken theory.
For longitudinal degrees
of freedom it is shown that the equivalence theorem can be used to obtain the correct double
logarithmic asymptotics. At the subleading level, corrections to the 
would be Goldstone bosons contribute which should be considered separately.
Explicit comparisons with existing one loop calculations are made.
\end{abstract}

\vskip12pt     
     
\setcounter{footnote}{0} \renewcommand{\thefootnote}{\arabic{footnote}}     
     
\vfill     
\clearpage     
\setcounter{page}{1} \pagestyle{plain}     
     
\section{Introduction}     

Future high energy experiments at the LHC and possibly a linear collider (TESLA or the NLC for instance),
will probe the full non-Abelian nature of the electroweak Standard Model (SM).
Thus one has to view the photon in particular as a particle with non-Abelian
character. At energies much larger than the weak scale, Sudakov logarithms, originating from vector boson
exchange, can lead to significant radiative corrections. The double logarithms (DL) can be of order
${\cal O}(20$ \%$)$ at one loop in the TeV range and a few \% at the two loop level. In addition, 
subleading corrections can also be significant, especially if the experimental accuracy is of the order of
${\cal O}(1$ \%$)$. 

As of this writing, there is no complete two loop calculation in the electroweak theory due to the
complexity of the number and nature of processes involved. It is thus of considerable interest to
investigate terms which are potentially large and which can be resummed to all orders. In Ref. \cite{flmm}
the leading DL corrections were calculated and found to exponentiate. The results were obtained by
using the infrared evolution equation method \cite{kl} calculated with the massless fields of the unbroken
theory. The equation has a different kernel depending on the value of the infrared cutoff.

There are, however, some important differences of the electroweak theory with respect to an unbroken 
gauge theory. Since the physical cutoff of the massive gauge bosons is the weak scale $M\equiv M_{\rm w}$,
pure virtual correction lead to physical cross sections depending on the infrared ``cutoff''.
Only the photon needs to be treated in
a semi-inclusive way. 
Additional complications arise due to the mixing involved to make the mass eigenstates and the fact
that at high energies, the longitudinal degrees of freedom are not suppressed.
Furthermore, since the asymptotic states are not group singlets, it is expected
that fully inclusive cross sections contain Bloch-Nordsieck violating electroweak corrections \cite{ccc}.

In this paper we extend the method of Ref. \cite{flmm} to the next to leading order
for the case of virtual corrections at high energies where we can
neglect particle masses. In addition, we show that the results of Ref. \cite{flmm} are also valid for
longitudinal degrees of freedom, which at first sight, is far from obvious. The connection between the
calculation performed in the massless theory and the longitudinal degrees of freedom is provided by
the Goldstone boson equivalence theorem. Another complication in comparison with the unbroken non-Abelian
case is the mixing of the mass-eigenstates. Especially for the external $Z$-boson and the photon states, 
the corresponding corrections in general 
don't factorize with respect to the original amplitude. We indicate how
the corrections are given to subleading level in terms of the fields of the unbroken theory.

Finally we compare our results with existing one-loop corrections in the high energy approximation 
for on-shell
$W$-pair production in electron-positron scattering.
Although this comparison constitutes a strong test of our approach it should be mentioned that it
would be extremely helpful to compare the results obtained in this approach with a general method
in terms of the physical fields. In this context a two loop DL-calculation would help clarify the
situation with contradicting results in the literature \cite{cc} and a subleading one loop approach
\cite{dp} would
give further support to the results presented in this work.

\section{Logarithmic corrections in non-Abelian theories} \label{sec:na}

In this section we are concerned with virtual double and single logarithmic corrections to scattering
amplitudes in massless non-Abelian theories
at fixed angle with all invariants large with respect to an infrared cutoff $\mu$,
i.e. $\mu^2 \ll s_{j,l}\equiv 2 p_jp_l \sim s$. It must
be emphasized that in high energy collider experiments there are also contributions depending on
angular variables (i.e. u/t etc.) which can be of genuine subleading nature \cite{bddms}. The philosophy
adopted here is that terms of the type $\log \frac{s}{\mu^2} \log \frac{u}{t}$ etc.
should be calculated exactly at least at the one loop level. For the higher order
terms below, we are only concerned with the $\log \frac{s}{\mu^2}$ behavior with 
$\mu^2\ll s \sim |t| \sim |u|$.
All mass terms are neglected, 
i.e. we assume $m_i < \mu$.

We begin by reviewing the general method for virtual corrections in the DL-approximation following the
approach of Ref. \cite{flmm}.

\subsection{Double logarithmic corrections} \label{sec:dl}    
     
Sudakov effects have been widely discussed for non-Abelian gauge theories,     
such as $SU(N)$ and can be calculated in various ways (see, for instance,     
\cite{NA}).    
A general method of finding the DL asymptotics     
(not only of the Sudakov type) is based on the infrared evolution equations     
describing the dependence of the amplitudes on the infrared cutoff $\mu $ of     
the virtual particle transverse momenta \cite{kl}. This cutoff plays the     
same role as the fictitious photon mass $\lambda $ in QED, but, unlike $\lambda $, it is not     
necessary that it vanishes and it may take an arbitrary value. It can be introduced in     
a gauge invariant way by working, for instance, in a finite phase space     
volume in the transverse direction with linear size $l\sim 1/\mu $. Instead     
of calculating asymptotics of particular Feynman diagrams and summing these     
asymptotics for a process with $n$ external lines it is convenient to     
extract the virtual particle with the smallest value of $|{%
\mbox{\boldmath     
$k$}_{\perp }}|$ in such a way, that the transverse momenta $|{%
\mbox{\boldmath $k$}_{\perp }^{\prime }}|$ of the other virtual particles     
are much bigger     
\begin{equation}     
\mbox{\boldmath $k$}_{\perp }^{\prime ^{2}}\gg {\mbox{\boldmath $k$}_{\perp     
}^{2}}\gg \mu ^{2}\;.     
\end{equation}     
For the other particles ${\mbox{\boldmath $k$}_{\perp }^{2}}$ plays the role of     
the initial infrared cut-off $\mu ^{2}$.     
     
In particular, the Sudakov DL corrections are related to the exchange of     
soft gauge bosons. For this case the integral over the 
momentum $k$ of the     
soft (i.e. $|k^0|\ll \sqrt{s}$) virtual boson with the smallest 
${\mbox{\boldmath $k$}}_\perp$ can be factored off, which leads to the 
following infrared evolution equation:     
\begin{eqnarray}     
{\cal M}(p_1,...,p_n;\mu^2) & = & {\cal M}_{\rm Born}(p_1,...,p_n) -\frac{i}{2}     
\frac{g^2_s}{(2\pi)^4} \sum_{j,l=1, j \neq l}^n \int_{s \gg \mbox{{\scriptsize \boldmath $k$}}^2_\perp   
\gg \mu^2} \frac{d^4k}{k^2+i \epsilon} \;\;     
\frac{p_jp_l}{(kp_j)(kp_l)}  \nonumber \\     
& & \times \; T^a(j) T^a(l) {\cal M} (p_1,...,p_n;{\mbox{\boldmath $k$}%
^2_\perp}) \,,  \label{eq:vem}     
\end{eqnarray}     
where the amplitude ${\cal M}(p_1,...,p_n;{\mbox{\boldmath $k$}^2_\perp})$     
on the right hand side is to be taken on the mass shell, but with the     
substituted infrared cutoff: $\mu^2 \longrightarrow {\mbox{\boldmath $k$}%
^2_\perp}$. The generator $T^a(l) (a=1,...,N)$ acts on the color indices of     
the particle with momentum $p_l$. The non-Abelian gauge coupling is $g$.     
In Eq. (\ref{eq:vem}), and below, $\mbox{\boldmath $k$}_\perp$ denotes the component
of the gauge boson momentum $\mbox{\boldmath $k$}$ transverse to the particle 
emitting this boson. 
It can be expressed in invariant form
as $\mbox{\boldmath $k$}^2_\perp \equiv \min ( (kp_l)(kp_j)/(p_lp_j))$ for all $j \neq l$.

The above factorization is related to a non-Abelian generalization of the     
Gribov theorem\footnote{The non-Abelian
generalization of Gribov's theorem is given in Ref. \cite{flmm}, together with a description of 
its essential content.} for the amplitude of the Bremsstrahlung of a photon with     
small transverse momentum ${\mbox{\boldmath $k$}_{\perp }}$ in high energy hadron scattering     
\cite{vg}.  
 
The form in which we present
Eq. (\ref{eq:vem}) corresponds to a covariant gauge 
for the gluon with 
momentum $k$. Formally this expression can be written in a gauge invariant 
way if we include in the sum the term with $j=l$ (which does not give 
a DL contribution). Indeed, in this case we can 
substitute $p_{i}p_{j}$ by $-p_{i}^{\mu }p_{j}^{\nu }d_{\mu \nu }(k)$, where 
the polarization matrices of the boson $d_{\mu \nu }(k)$ in the various     
gauges differ by the terms proportional to $k^{\mu }$ or $k^{\nu }$ giving a
vanishing contribution due to the conservation of the total color charge $
\sum_{a}T^{a}=0$. Thus we have the possibility of choosing appropriate     
gauges for each kinematical region of quasi-collinearity of $k$ and     
$p_{l}$. We can, however, use (\ref{eq:vem}) as well, noting that in this region
for $j\neq l$ we have $p_{j}p_{l}/kp_{j}\simeq E_{l}/\omega $, where $E_{l}$
is the energy of the particle with momentum $p_{l}$ and $\omega$ the
frequency of the emitted gauge boson, so that:     
\begin{eqnarray}
{\cal M}(p_{1},...,p_{n};\mu ^{2}) &=&{\cal M}_{\rm Born}(p_{1},...,p_{n})-\frac{%
2g^{2}_s}{(4\pi )^{2}}\sum_{l=1}^{n}\int_{\mu ^{2}}^{s}\frac{d{%
\mbox{\boldmath $k$}_{\perp }^{2}}}{{\mbox{\boldmath $k$}_{\perp   
}^{2}}}\int_{|\mbox{\scriptsize \boldmath $k$}_{\perp }|/\sqrt{s}}^1   
\frac{dv}{v}  \nonumber \\     
&&\times \;C_{l}{\cal M}(p_{1},...,p_{n};{\mbox{\boldmath     
$k$}_{\perp }^{2}}%
)\;\;,  \label{eq:dgvem}     
\end{eqnarray}     
where $C_{l}$ is the eigenvalue of the Casimir operator $T^{a}(l)T^{a}(l)$     
($%
C_{l}=C_{A}$ for gauge bosons in the adjoint representation of the gauge     
group $SU(N)$ and $C_{l}=C_{F}$ for fermions in the fundamental     
representation).     
In this last step we also used the identity $\sum_{j=1}^n T^a(j) {\cal M}(p_1,...,p_n;{\mbox{\boldmath
$k$}_{\perp }^{2}})=0$, corresponding to the conservation of the total group charge.
The integral over $d^4k$ was written in terms of the Sudakov components according to the discussion
in section \ref{sec:sd} upon replacing the longitudinal component $u$ with the boson on-shell
expression $s u v =\mu^2+{\mbox{\boldmath $k$}_{\perp }^{2}}$.
Thus, in Sudakov DL corrections there are no interference effects, so that we can talk about
the emission (and absorption) of a gauge boson by a definite (external) particle, namely by a particle
with momentum almost collinear to ${\mbox{\boldmath $k$}}$.
     
The differential form of the infrared evolution equation follows immediately     
from (\ref{eq:dgvem}):     
\begin{equation}     
\frac{\partial {\cal M}(p_{1},...,p_{n};\mu ^{2})}{\partial \log (\mu     
^{2})}%
=K(\mu ^{2}){\cal M}(p_{1},...,p_{n};\mu ^{2})\,,  \label{eq:ee}     
\end{equation}     
where     
\begin{equation}     
K(\mu ^{2})\equiv -\frac{1}{2}\sum_{l=1}^{n}\frac{\partial W_{l}(s,\mu     
^{2})%
}{\partial \log (\mu ^{2})}     
\end{equation}     
with     
\begin{equation}     
W_{l}(s,\mu ^{2})=\frac{g^{2}_s}{(4\pi)^{2}}C_{l}\, \log^2 \frac{s}{\mu^2}    
\,.  \label{eq:wl}     
\end{equation}     
$W_l$ is the probability to emit a soft and     
almost collinear gauge boson from the particle $l$,      
subject to the infrared cut-off $\mu $ on the transverse momentum \cite{flmm}. Note  
again that the cut-off $\mu $ is not taken to zero.     
To logarithmic accuracy, we obtain directly from (\ref{eq:wl}):     
\begin{equation}     
\frac{\partial W_{l}(s,\mu ^{2})}{\partial \log (\mu ^{2})}=-\frac{g^{2}_s}{%
8\pi ^{2}}C_{l}\log \frac{s}{\mu^2}\,.     
\end{equation}     
The infrared evolution equation (\ref{eq:ee}) should be solved with an     
appropriate initial condition. In the case of large scattering angles, if we    
choose the cut-off to be the large scale $s$ then clearly there are no     
Sudakov corrections. The initial condition is therefore     
\begin{equation}     
{\cal M}(p_{1},...,p_{n};s)={\cal M}_{\rm Born}(p_{1},...,p_{n}),     
\end{equation}     
and the solution of (\ref{eq:ee}) is thus given by the product of the Born     
amplitude and the Sudakov form factors:     
\begin{equation}     
{\cal M}(p_{1},...,p_{n};\mu ^{2})={\cal M}_{\rm Born}(p_{1},...,p_{n})\exp     
\left( -\frac{1}{2}\sum_{l=1}^{n}W_{l}(s,\mu ^{2})\right)     
\end{equation}     
Therefore we obtain an exactly analogous Sudakov exponentiation for the     
gauge group $SU(N)$ to that for the Abelian case \cite{s}.     

\subsection{Soft divergences in the massless theory} \label{sec:sd}     

In this section we briefly review the types of soft, i.e. $|k^0| \ll \sqrt{s}$, divergences in loop
corrections with massless particles. In general, those contributions, unlike the collinear 
logarithms, can be obtained by setting all $k$ dependent terms in the numerator of tensor
integrals to zero (since the terms left are of the order of the hard scale $s$). 
Thus it is
clear that the tensor structure which emerges is that of the inner scattering amplitude in Fig.
\ref{fig:nll} taken on the mass-shell, times a scalar function of the given loop correction. 
In the Feynman gauge, for instance, we find for the
well known vertex corrections the familiar three-point function $C_0$ and for higher point
functions we note that in the considered case all infrared divergent scalar integrals reduce to
$C_0$ multiplied by factors of $\frac{1}{s}$ etc.. The only infrared divergent three point function
is given by
\begin{equation}
C_0(s/\mu^2) \equiv \int_{{\mbox{\boldmath $k$}^2_{\perp}} > \mu^2} \frac{d^4k}{(2\pi)^4} \frac{1}{
(k^2+i\varepsilon)(k^2+2p_jk+i\varepsilon)(k^2-2p_lk+i\varepsilon)}
\label{eq:c0mudef}
\end{equation}
It is now convenient to use the Sudakov parametrization for the exchanged virtual boson:
\begin{equation}
k=vp_j \; + u p_l + k_\perp \label{eq:sudp}
\end{equation}
For the boson propagator we use the identity
\begin{equation}
\frac{i}{s u v - {\mbox{\boldmath $k$}^2_{\perp}}+i\varepsilon}= {\cal P}
\frac{i}{s u v - {\mbox{\boldmath $k$}^2_{\perp}}} + \pi \delta ( s u v - {\mbox{\boldmath $k$}^2_{
\perp}}) \label{eq:propid}
\end{equation}
writing it in form of the real and imaginary parts (the principle value is indicated by ${\cal P}$).
The latter does not contribute to the DL asymptotics and at higher orders gives subsubleading 
contributions. 
Rewriting the measure as $d^4k=d^2k_\perp d^2k_\parallel$ with
\begin{eqnarray}
d^2k_\perp &=& |{\mbox{\boldmath $k$}_{\perp}}| d |{\mbox{\boldmath $k$}_{\perp}}| d \phi =
\frac{1}{2} d {\mbox{\boldmath $k$}^2_{\perp}} d \phi = \pi d {\mbox{\boldmath $k$}^2_{\perp}}
\\
d^2k_\parallel &=& | \partial (k^0,k^x)/ \partial (u,v)| d u d v \approx \frac{s}{2} du dv
\end{eqnarray}
where we turn the coordinate system such that the $p_j,p_l$ plane corresponds to $0,x$ and the
$y,z$ coordinates to the $k_\perp$ direction so that it is purely spacelike. 
The last equation follows from $p_l^2=0$, i.e. $p_{l_x}^2\approx p_{l_0}^2$ and 
\begin{equation}
(p_{j_0}p_{l_x}-p_{l_0}p_{j_x})^2 \approx (p_{j_0}p_{l_0}-p_{l_x}p_{j_x})^2=(p_jp_l)^2=s/2
\end{equation}
The function $C_0(s/\mu^2)$ is fastly converging for large ${\mbox{\boldmath $k$}^2_{\perp}}$ and we
are interested here in the region $\mu^2 \ll s$ in order to obtain large logarithms. Then
logarithmic corrections come from the region ${\mbox{\boldmath $k$}^2_{\perp}}\ll s|u|, s|v| \ll s$
(the strong inequalities give DL, the simple inequalities single ones)
and we can write to logarithmic accuracy:
\begin{eqnarray}
C_0(s/\mu^2) &=& \frac{s \pi}{2 (2 \pi)^4} \int^\infty_{-\infty}du \int^\infty_{-\infty} dv 
\int^\infty_{\mu^2} d {\mbox{\boldmath 
$k$}^2_{\perp}} \times \nonumber \\ && 
\frac{1}{(s u v - {\mbox{\boldmath $k$}^2_{\perp}}+i\varepsilon)(
s u v - {\mbox{\boldmath $k$}^2_{\perp}}+su +i\varepsilon) (
s u v - {\mbox{\boldmath $k$}^2_{\perp}}-s v +i\varepsilon)} \nonumber \\
&\approx& \frac{s i \pi^2}{2 (2 \pi)^4} \int^\infty_{-\infty} \frac{du}{su} \int^\infty_{-\infty} 
\frac{dv}{sv} \int^\infty_{-\infty} 
d {\mbox{\boldmath$k$}^2_{\perp}} \Theta ( {\mbox{\boldmath$k$}^2_{\perp}}-
\mu^2 ) \delta ( suv-{\mbox{\boldmath$k$}}^2_{\perp}) \nonumber \\
&\approx&  \frac{i}{2(4\pi)^2s} \int^{1}_{-1} \frac{du}{u} \int^{1}_{-1} \frac{dv}{v} \Theta (suv - \mu^2)
\nonumber \\
&=&  \frac{i}{(4\pi)^2s} \int^{1}_{0} \frac{du}{u} \int^{1}_{0} \frac{dv}{v} \Theta (suv - \mu^2)
\nonumber \\
&=&  \frac{i}{(4\pi)^2s} \int^1_\frac{\mu^2}{s} \frac{du}{u} \int^1_\frac{\mu^2}{su} \frac{dv}{v}
\nonumber \\
&=&  \frac{i}{2(4\pi)^2s} \log^2 \frac{s}{\mu^2}
\end{eqnarray}
Thus, no single soft logarithmic corrections are present in $C_0(s/\mu^2)$.
In order to see that this result is not just a consequence of our regulator, we repeat the calculation
for a fictitious gluon mass\footnote{Note that this regulator spoils gauge invariance and leads to
possible inconsistencies at higher orders. Great care must be taken for instance when a three gluon
vertex is regulated inside a loop integral.}. In this case we have
\begin{equation}
C_0(s/\lambda^2) \equiv \int \frac{d^4k}{(2\pi)^4} \frac{1}{(k^2-\lambda^2+i\varepsilon)
(k^2+2p_jk+i\varepsilon)(k^2-2p_lk+i\varepsilon)}
\label{eq:c0def}
\end{equation}
It is clear that $C_0(s/\lambda^2)$ contains soft and collinear divergences ($k \parallel p_{j,l}$) and
is regulated with the cutoff $\lambda$, which plays the role of $\mu$ in this case.
Integrating over Feynman parameters we find:
\begin{equation}
C_0(s/\lambda^2)=\frac{i}{(4\pi)^2s} \left( \frac{1}{2} \log^2 \frac{s}{-\lambda^2+i \varepsilon} + \frac{\pi^2}{3}
\right) \label{eq:c0res}
\end{equation}
We are only interested here in the real part of loop corrections of scattering
amplitudes since they are multiplied by the Born amplitude and the imaginary pieces contribute
to cross sections at the next to next to leading level as mentioned above. In fact, the minus sign
inside the double logarithm corresponds precisely to the omitted principle value contribution
of Eq. (\ref{eq:propid}) in the previous calculation.
Thus, no single soft logarithmic correction is present in the case when particle masses can be
neglected. 

This feature prevails to higher orders as well since it has been shown that also in non-Abelian 
gauge theories
the one-loop Sudakov form factor exponentiates \cite{NA}.

In case we would keep mass-terms, even two point functions, which in our scheme can only
yield collinear logarithms, would contain a soft logarithm due to the mass-renormalization which
introduces a derivative contribution \cite{m}. In conclusion, all leading
soft corrections are contained in double logarithms (soft and collinear)
and subleading logarithmic corrections
in a massless theory, with all invariants large ($s_{j,l}=2p_jp_l\sim {\cal O}(s)$) compared to the
infrared cutoff,
are of the collinear type or renormalization group logarithms. 

\subsection{Virtual logarithmic corrections from the Altarelli-Parisi splitting functions} \label{sec:ap}

\begin{figure} 
\centering 
\epsfig{file=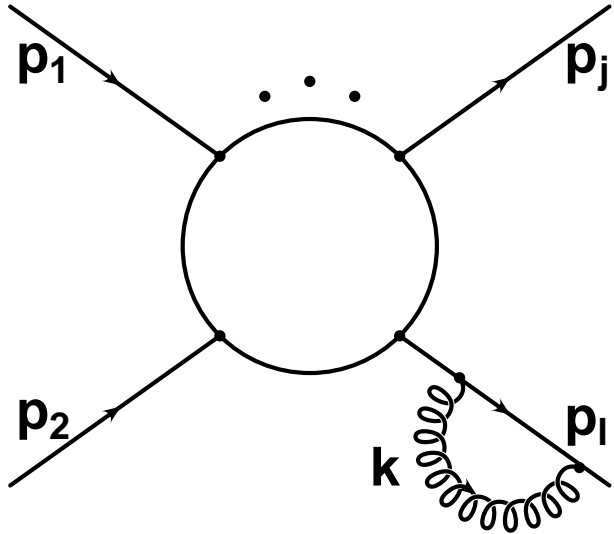,width=8cm} 
\caption{In an axial gauge, all collinear logarithms come from corrections to a particular external line 
(depending on the choice of the four vector $n^\nu$ satisfying $n^\nu A^a_\nu=0$) as
illustrated in the figure. In a covariant gauge, the sum over all possible
insertions is reduced to a sum over all $n$-external legs due to Ward identities. Overall, these
corrections factorize with respect to the Born amplitude.}
\label{fig:col}
\end{figure}

In an axial gauge, collinear logarithms are related to corrections
on a particular external leg depending on the choice of the four vector $n_\nu$ \cite{col}. 
A typical diagram is depicted in Fig. \ref{fig:col}. In a general covariant
gauge this corresponds (using Ward identities)
to a sum over insertions in all $n$ external legs \cite{flmm}.
We can therefore adopt the strategy to extract the gauge invariant contribution from the external
line corrections on the invariant matrix element at the subleading level.
The results of the previous section are thus 
important in that they allow the use of the Altarelli-Parisi
approach to calculate the subleading contribution to the evolution kernel of Eq. (\ref{eq:ee}).
We are here only concerned with virtual corrections and use the universality of the splitting
functions to calculate the subleading terms. For this purpose we use the virtual quark 
and gluon contributions to the splitting functions
$P^V_{qq}(z)$ and $P^V_{gg}(z)$ describing the probability to emit a soft and/or collinear virtual
particle with energy fraction $z$ of the original external line four momentum.
The infinite momentum frame corresponds to the Sudakov parametrization with lightlike vectors.
In general, the splitting functions $P_{BA}$ 
describe the probability of finding a particle $B$ inside a particle $A$
with fraction $z$
of the longitudinal momentum of $A$ with probability
${\cal P}_{BA}$ to first order \cite{ap}:
\begin{equation}
d {\cal P}_{BA}(z)=\frac{\alpha_s}{2\pi} P_{BA} d t
\end{equation}
where the variable $t=\log \frac{s}{\mu^2}$ for our purposes. It then follows \cite{ap} that
\begin{equation}
d {\cal P}_{BA}(z)=\frac{\alpha_s}{2\pi} \frac{z(1-z)}{2} \overline{\sum_{spins}} \frac{|V_{A
\longrightarrow B+C}|^2}{{\mbox{\boldmath$k$}^2_{\perp}}} d \log {\mbox{\boldmath$k$}^2_{\perp}} \label{eq:dpabres}
\end{equation}
where $V_{A\longrightarrow B+C}$ denotes the elementary vertices and
\begin{equation}
P_{BA}(z)=\frac{z(1-z)}{2} \overline{\sum_{spins}} \frac{|V_{A
\longrightarrow B+C}|^2}{{\mbox{\boldmath$k$}^2_{\perp}}} 
\end{equation}
The upper bound on the integral over $d {\mbox{\boldmath$k$}^2_{\perp}}$ in Eq. (\ref{eq:dpabres}) 
is $s$ and it is thus
directly related to $d t$.
Regulating the virtual infrared divergences with the transverse momentum cutoff as described above,
we find the virtual contributions to the splitting functions for external quark and gluon lines:
\begin{eqnarray}
P^V_{qq}(z)&=& C_F \left( - 2 \log \frac{s}{\mu^2} + 3 \right) \delta(1-z) \label{eq:pqqv} \\
P^V_{gg}(z)&=& C_A \left( - 2 \log \frac{s}{\mu^2} + \frac{4}{C_A} \beta^{\rm QCD}_0 \right) \delta(1-z) \label{eq:pggv} 
\end{eqnarray}
The functions can be calculated directly from loop corrections to the elementary
processes \cite{aem,a,dot} and the logarithmic term corresponds to the
leading kernel of section \ref{sec:dl}. 
We introduce virtual distribution functions which include only the effects of loop computations.
These fulfill the Altarelli-Parisi equations\footnote{Note that the off diagonal
splitting functions $P_{qg}$ and $P_{gq}$ do not contribute to the virtual probabilities to the order
we are working here. In fact, for virtual corrections there is no need to introduce off-diagonal terms
as the corrections factorize with respect to the Born amplitude. The normalization of the Eqs.
(\ref{eq:pqqv}) and (\ref{eq:pggv}) corresponds to calculations in two to two processes on the cross section
level with the gluon symmetry factor $\frac{1}{2}$ included. The results, properly normalized,
are process independent.}
\begin{eqnarray}
\frac{\partial q(z,t)}{\partial t}&=& \frac{\alpha_s}{2\pi} \int^1_z \frac{dy}{y} q(z/y,t) P^V_{qq}(y)
\label{eq:apqq} \\
\frac{\partial g(z,t)}{\partial t}&=& \frac{\alpha_s}{2\pi} \int^1_z \frac{dy}{y} g(z/y,t) P^V_{gg}(y)
\label{eq:apgg} 
\end{eqnarray}
The splitting functions are related by
$P_{BA}=P^R_{BA}+P^V_{BA}$, where
$R$ denotes the contribution from real gauge boson emission\footnote{$P_{qq}$ was first calculated by
V.N.~Gribov and L.N.~Lipatov in the context of QED \cite{gl}.}. $P_{BA}$ is free of logarithmic
corrections and positive definite.
The subleading term in Eq. (\ref{eq:pggv})  
indicates that the only subleading corrections in the pure
glue sector are related to a shift in the scale of the coupling. These corrections enter with a
different sign compared to the conventional running coupling effects.
The renormalizations with respect to the Born amplitude
as well as the ones belonging to the
next to leading terms at higher orders will be indicated below by
writing $\alpha_s(\overline{\mu}^2)
\longrightarrow \alpha_s(s)$. 
For fermion lines there is an additional
subleading correction from collinear terms which is not related to a change in the scale of the 
coupling.

Inserting the virtual probabilities of Eqs. (\ref{eq:pqqv}) and (\ref{eq:pggv}) into the Eqs.
(\ref{eq:apqq}) and (\ref{eq:apgg}) we find:
\begin{eqnarray}
q(1,t)&=&q_0 \exp \left[ - \frac{ \alpha_s(s) C_F}{2 \pi} \left( \log^2 \frac{s}{\mu^2}
- 3 \log \frac{s}{\mu^2} \right) \right] \label{eq:qsol} \\
g(1,t)&=&g_0 \exp \left[ - \frac{ \alpha_s(s) C_A}{2 \pi} \left( \log^2 \frac{s}{\mu^2}
- \frac{4}{C_A} \beta^{\rm QCD}_0 \log \frac{s}{\mu^2} \right) \right] \label{eq:gsol} 
\end{eqnarray}
where $\beta^{\rm QCD}_0=\frac{11}{12}C_A-\frac{1}{3}T_Fn_f$ with $C_A=3$ and $T_F=\frac{1}{2}$.
\begin{figure} 
\centering 
\epsfig{file=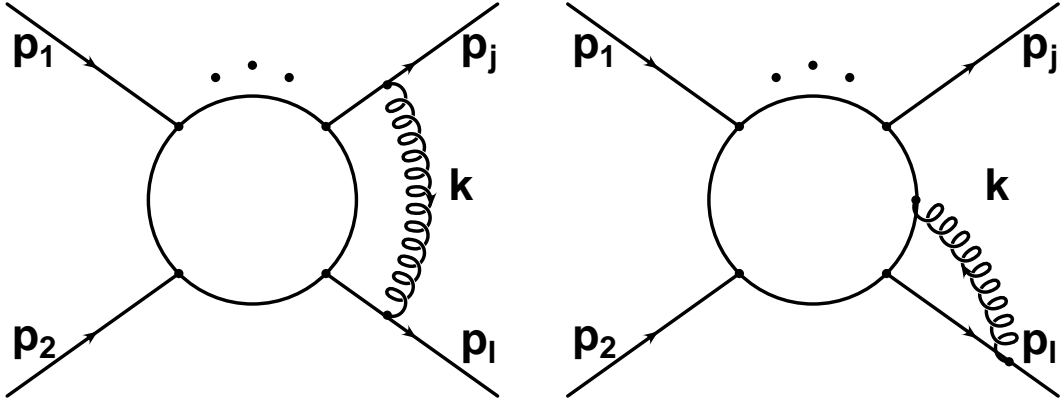,width=14cm} 
\caption{Feynman diagrams contributing to the infrared evolution 
equation (\ref{eq:mrg}) for a process with $n$ external legs. In a general
covariant gauge the
virtual gluon with the smallest value of ${\mbox{\boldmath $k$}}_{\perp}$ is attached to 
different external lines. The inner scattering amplitude is assumed to be 
on the mass shell.} \label{fig:nll}
\end{figure} 
These functions describe the total contribution for the emission of virtual particles (i.e. $z=1$), 
with all invariants large compared to the cutoff $\mu$, to the
densities $q(z,t)$ and $g(z,t)$. The normalization is on the level of the cross section.
For the invariant matrix element we thus find at the subleading level:
\begin{eqnarray}
{\cal M} (p_1,...,p_n,g_s,\mu)&=&{\cal M} (p_1,...,p_n,g_s(s)) \times \nonumber \\
&& \exp \left(- \frac{1}{2}
\sum^{n_q}_{j=1} W^q_j(s,\mu^2)-\frac{1}{2} \sum^{n_g}_{l=1} W^g_l(s,\mu^2) \right)
\label{eq:msol}
\end{eqnarray}
with $n_q+n_g=n$, and 
\begin{eqnarray}
W^q(s,\mu^2)&=&\frac{ \alpha_s(s) C_F}{4 \pi} \left( \log^2 \frac{s}{\mu^2}
- 3 \log \frac{s}{\mu^2} \right) \label{eq:wq} \\
W^g(s,\mu^2)&=&\frac{ \alpha_s(s) C_A}{4 \pi} \left( \log^2 \frac{s}{\mu^2} - \frac{4}{C_A} 
\beta^{\rm QCD}_0 \log \frac{s}{\mu^2} \right)
\label{eq:wg} 
\end{eqnarray}
Again we note that the running coupling notation in the Born-amplitude of Eq. (\ref{eq:msol}) denotes
the renormalization corrections of the Born amplitude and higher order corrections.
The functions $W^q$, $W^g$ correspond to the probability of emitting a virtual 
soft and/or collinear gauge
boson from the particle $q$, $g$ subject to the infrared cutoff $\mu$. Typical diagrams contributing
to Eq. (\ref{eq:msol}) in a covariant gauge are depicted in Fig. \ref{fig:nll}. 
In massless QCD there is no need for the label $W^q_j$ or $W^g_l$, however, we write it for later convenience.
The universality of the splitting functions is crucial in obtaining the above result.

\subsection{Renormalization group equation} \label{sec:rg}

\begin{figure} 
\centering 
\epsfig{file=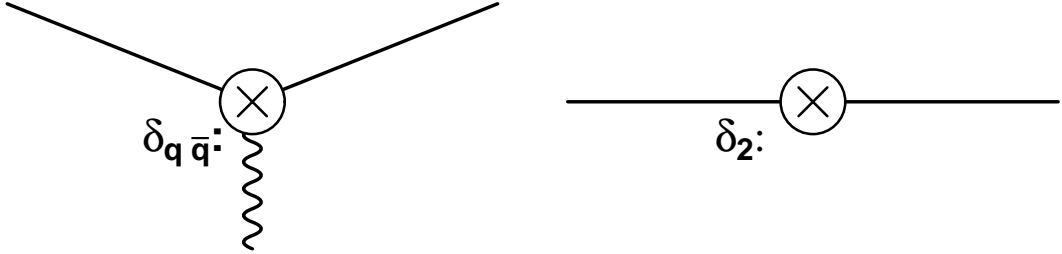,width=14cm} 
\caption{The two counterterms contributing to the quark anomalous dimension $\gamma_{q\overline{q}} 
= \frac{\partial}{\partial \log \overline{\mu}^2} \left( -\delta_{q \overline{q}}+\delta_2 \right)$.
Here $\overline{\mu}$ denotes the $\overline{\rm MS}$ dimensional regularization mass parameter.
Due to divergences in loop corrections there are scaling violations also in the massless theory.}
\label{fig:gqq}
\end{figure}

The solution presented in Eq. (\ref{eq:msol}) determines the evolution of the virtual scattering
amplitude ${\cal M} (p_1,...,p_n,g_s,\mu)$ for large energies at fixed angles and subject to
the infrared regulator $\mu$. In the massless case there is a one to one correspondence between the
high energy limit and the infrared limit as only the ratio $s/\mu^2$ enters as a dimensionless 
variable \cite{pqz,p}.
Thus, we can generalize the Altarelli-Parisi equations (\ref{eq:apqq}) and (\ref{eq:apgg}) to the invariant
matrix element in the language of the renormalization group. For this purpose, we define the
infrared singular (logarithmic) anomalous dimensions 
\begin{equation}
\Gamma_q (t) \equiv \frac{C_F \alpha_s}{4 \pi} t \;\;;\;\; \Gamma_g (t) \equiv \frac{C_A \alpha_s}{4 \pi} t
\label{eq:irad}
\end{equation}
Infrared divergent anomalous dimensions have been derived in the context of renormalization properties
of gauge invariant Wilson loop functionals \cite{kr}. In this context they are related to undifferentiable
cusps of the path integration and the cusp angle $~p_jp_l/\mu^2$ gives rise to the logarithmic nature
of the anomalous dimension. In case we use off-shell amplitudes, one also has contributions from end
points of the integration \cite{kr}. 
The leading terms in the equation below have also been discussed in Refs. \cite{ct}, \cite{ar} and
\cite{ku} in the context of QCD.
With these notations we
find that Eq. (\ref{eq:msol}) satisfies
\begin{eqnarray}
&& \left( \frac{\partial}{\partial t} + \beta^{\rm QCD} \frac{\partial}{\partial g_s} + n_g \left(
\Gamma_g(t)- \frac{1}{2} \frac{\alpha_s}{\pi} \beta^{\rm QCD}_0 \right) + n_q 
\left( \Gamma_q(t) + \frac{1}{2} \gamma_{q\overline{q}} \right) \right) \nonumber \\ 
&& \times {\cal M} (p_1,...,p_n,g_s,\mu) =0 \label{eq:mrg}
\end{eqnarray}
to the order we are working here and 
where ${\cal M}(p_1,...,p_n,g_s,\mu)$ is taken on the mass-shell.
The difference in the sign of the derivative term compared to Eq. (\ref{eq:ee}) is due to the fact
that instead of differentiating with respect to $\log \mu^2$ we use $\log s/\mu^2$.
The quark-antiquark operator anomalous dimension $\gamma_{q\overline{q}}=-
C_F \frac{3}{4} \frac{\alpha}{\pi}$ enters even for massless theories as the quark antiquark
operator leads to
scaling violations through loop effects since the quark masslessness is not protected by gauge
invariance and a dimensionful infrared cutoff needs to be introduced. Thus, although the Lagrangian
contains no $m q\overline{q}$ term, quantum corrections lead to the anomalous scaling violations
in the form of $\gamma_{q\overline{q}}$. The factor $\frac{1}{2}$ occurs since we write Eq. (\ref{eq:mrg})
in terms of each external line
separately\footnote{In case of a massive theory, we could, for instance avoid the
anomalous dimension term $\gamma_{q\overline{q}}$ by adopting the pole mass definition. In this case, however, we
would obtain terms in the wave function renormalization, and in any case, the one to one correspondence
between UV and IR scaling, crucial for the validity of Eq. (\ref{eq:mrg}), is violated.}.
For the gluon, the scaling violation due to the infrared cutoff are manifest in 
terms of an anomalous dimension proportional to the $\beta$-function since the
gluon mass is protected by gauge invariance from loop corrections. 
Thus, in the bosonic sector
the subleading terms correspond effectively to a scale change of the  
coupling.
Fig. \ref{fig:gqq} illustrates the corrections to the external quark-antiquark lines from loop effects.

Except for the infrared singular anomalous dimension (Eq. (\ref{eq:irad})), all other terms 
in Eq. (\ref{eq:mrg}) are the standard contributions to the renormalization group equation for
S-matrix elements \cite{sterm}. In QCD, observables with infrared singular anomalous dimensions, regulated
with a fictitious gluon mass, are ill
defined due to the masslessness of gluons. 
In the electroweak theory, however, we can legitimately
investigate only virtual corrections since the gauge bosons will require a mass. 
Eq. (\ref{eq:mrg}) will thus be very useful in the following sections.

\section{Logarithmic corrections in broken gauge theories} \label{sec:bg}

In the following we will apply the results obtained in the previous sections to the case of spontaneously
broken gauge theories. It will be necessary, at least at the subleading level, to distinguish between
transverse and longitudinal degrees of freedom. The physical motivation in this approach is that for
very large energies, $s \gg M_W^2 \equiv M^2$, the electroweak theory is in the unbroken phase, with an
exact $SU(2) \times U(1)$ gauge symmetry. We will calculate the corrections to this theory and use
the high energy solution as a matching condition for the regime for values of $\mu < M$.

We begin by considering some simple kinematic arguments for massive vector bosons.
A vector boson at rest has momentum $k^\nu=(M,0,0,0)$ and a polarization vector that is a linear
combination of the three orthogonal unit vectors
\begin{equation}
e_1\equiv (0,1,0,0)\;\;\;,\;\;\; e_1\equiv (0,0,1,0)\;\;\;,\;\;\;e_3\equiv (0,0,0,1) \;\;\;. \label{eq:edef}
\end{equation}
After boosting this particle along the $3$-axis, its momentum will be $k^\nu=(E_k,0,0,k)$. The three
possible polarization vectors are now still satisfying:
\begin{equation}
k_\nu e^\nu_j = 0 \;\;\;,\;\;\; e_j^2=-1 \;\;\;. \label{eq:econd}
\end{equation}
Two of these vectors correspond to $e_1$ and $e_2$ and describe the transverse polarizations. The
third vector satisfying (\ref{eq:econd}) is the longitudinal polarization vector
\begin{equation}
e_L^\nu (k) = (k/M,0,0,E_k/M) \label{eq:lpv}
\end{equation}
i.e. $e_L^\nu (k)=k^\nu/M + {\cal O}(M/E_k)$ for large energies. These considerations illustrate that
the transversely polarized degrees of freedom at high energies are related to the massless theory,
while the longitudinal degrees of freedom need to be considered separately.

Another manifestation of the different high energy nature of the two polarization states is
contained in the Goldstone boson equivalence theorem.
It states that the unphysical Goldstone boson that is ``eaten
up'' by a massive gauge boson still controls its high energy asymptotics. A more precise formulation
is given below in section \ref{sec:ld}.

Thus we can legitimately use the results obtained in the massless non-Abelian theory if we restrict ourselves to
the transverse degrees of freedom at high energies.
We will, however, show
that to DL accuracy the results of Ref. \cite{flmm} can be used in connection with the Goldstone boson
equivalence theorem.

Another difference to the situation in an unbroken non-Abelian theory is the mixing of the physical fields
with the fields in the unbroken phase. These complications are especially relevant for the $Z$-boson
and the photon.

\subsection{Results for transverse degrees of freedom} \label{sec:td}

\begin{figure} 
\centering 
\epsfig{file=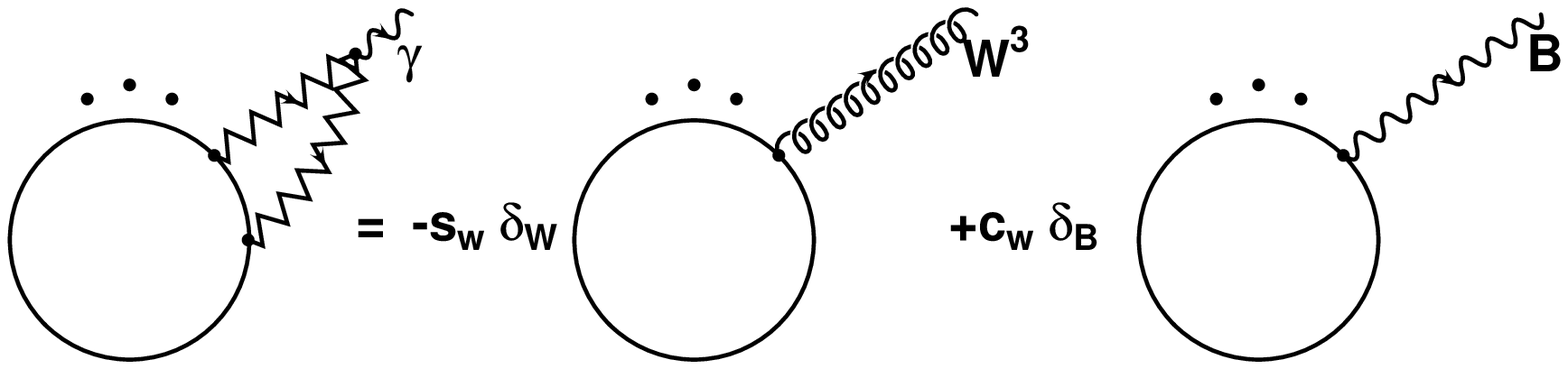,width=14cm} 
\vspace{0.5cm} \\ 
\epsfig{file=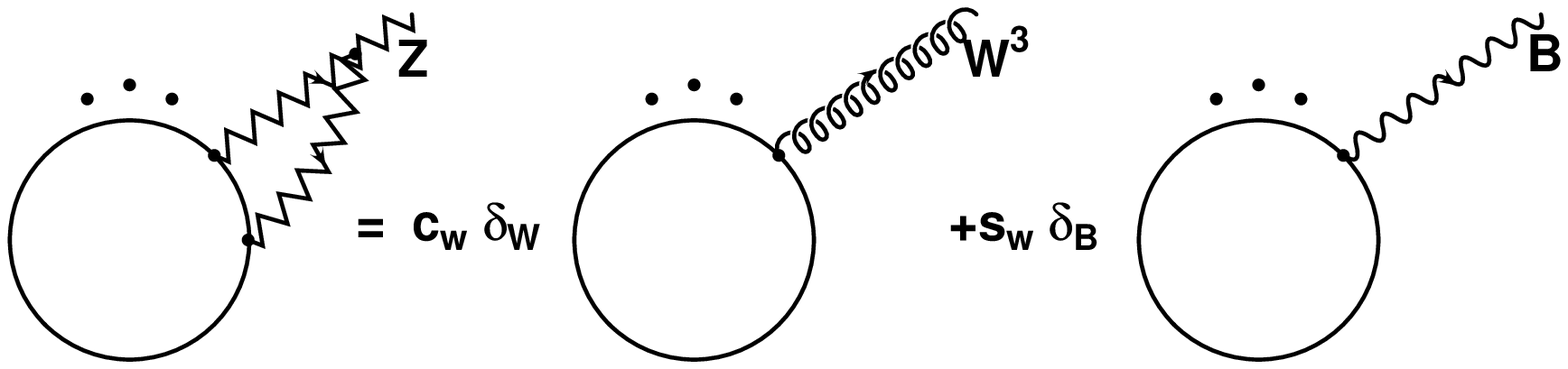,width=14cm} 
\vspace{0.5cm} \\ 
\epsfig{file=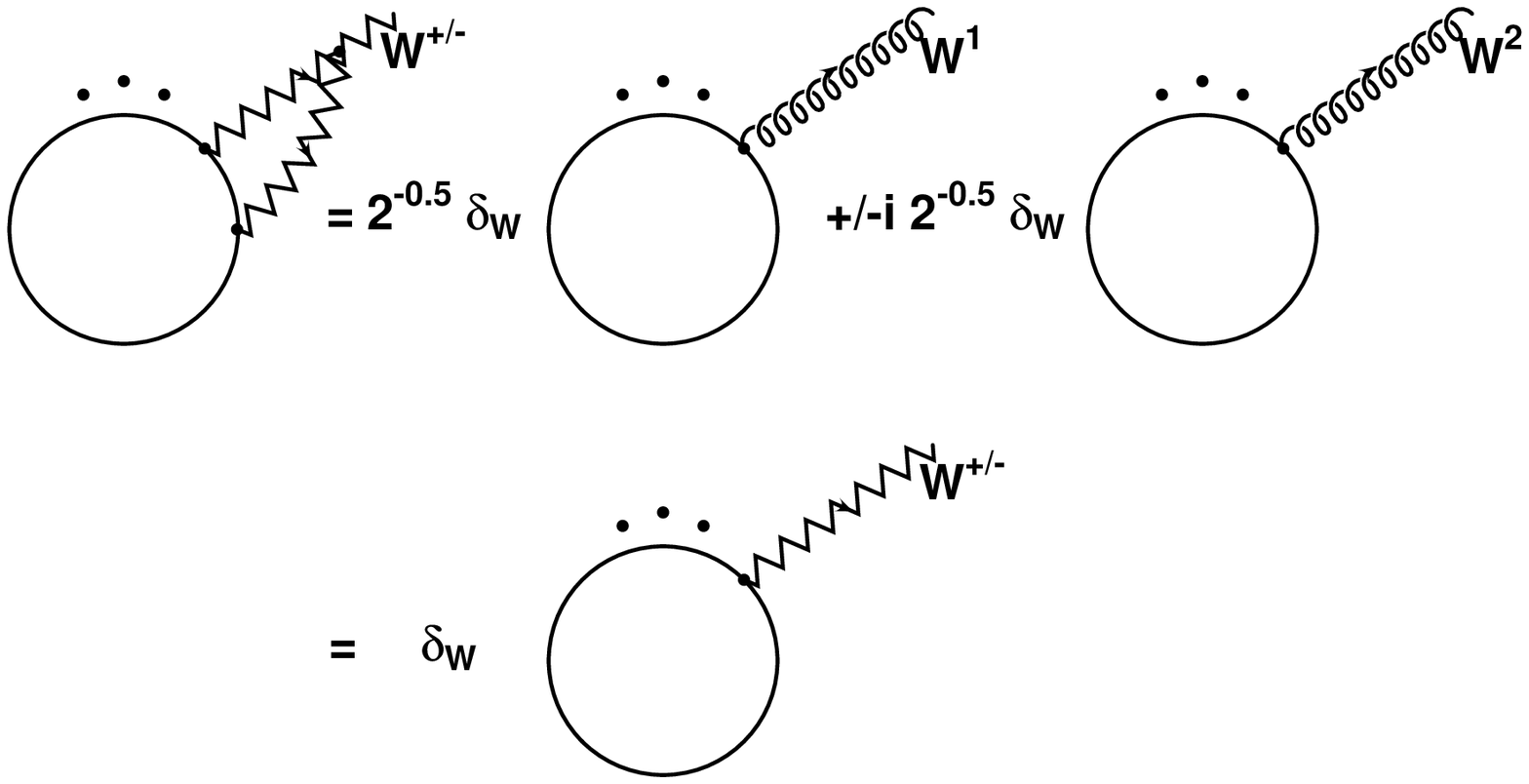,width=14cm} 
\caption{The schematic corrections to external gauge boson emissions in terms of the fields in the unbroken phase
of the electroweak theory. There are no mixing terms between the $W^3_\nu$ and $B_\nu$ fields
for massless fermions. We denote $\cos \theta_{\rm w}$ by $c_{\rm w}$ and $\sin \theta_{\rm w}$ by 
$s_{\rm w}$.
For $W^\pm$ final states, the corrections factorize with respect to the
physical amplitude. In general, one has to sum over all fields of the unbroken theory with each
amplitude being multiplied by the respective mixing coefficient.}
\label{fig:mix}
\end{figure}

The results we obtain in this section are generally valid for spontaneously
broken gauge theories, however, for
definiteness we discuss only the electroweak Standard Model. The physical gauge bosons are thus a 
massless photon (described by the field $A_\nu$) and massive $W^\pm$ and $Z$ bosons
(described correspondingly by fields $W_{\nu }^{\pm }$ and     
$Z_{\nu }$).:
\begin{eqnarray}
W^\pm_\nu&=& \frac{1}{\sqrt{2}} \left( W^1_\nu \pm i W^2_\nu \right) \label{eq:wpm} \\ 
Z_\nu&=& \cos \theta_{\rm w}W^3_\nu + \sin \theta_{\rm w} B_\nu \label{eq:z} \\
A_\nu&=& -\sin \theta_{\rm w}W^3_\nu + \cos \theta_{\rm w} B_\nu \label{eq:p} 
\end{eqnarray}
Thus, amplitudes containing physical fields will correspond to a linear combination of the
massless fields in the unbroken phase. The situation is illustrated schematically for a single
gauge boson external leg in Fig. \ref{fig:mix}. In case of the $W^\pm$ bosons, the corrections 
factorize with respect to the physical amplitude.

To logarithmic accuracy, all masses can be set equal:     
\[     
M_{Z}\sim M_{W}\sim M_{\rm Higgs}\sim M     
\]     
and the energy considered to be much larger, $\sqrt{s}\gg M$.      
The physical fields are given in terms of the unbroken fields according to Eqs. (\ref{eq:wpm}), (\ref{eq:z})
and (\ref{eq:p}).
The left and right handed     
fermions are correspondingly doublets ($T=1/2$) and singlets ($T=0$) of the     
$%
SU$(2) weak isospin group and have hypercharge $Y$ related to the electric     
charge $Q$, measured in units of the proton charge, by the     
Gell-Mann-Nishijima formula $Q=T^{3}+Y/2$.     

The value for the infrared cutoff $\mu$ can be chosen in two different regimes:
1) $\sqrt{s} \gg \mu \gg M$ and 2) $\mu \ll M$. The second case is universal in the sense
that it does not depend on the details of the electroweak theory and will be discussed below.
In the first region we can neglect spontaneous symmetry breaking effects (in particular
gauge boson masses) and consider the theory with fields $B_{\nu }$ and $W_{\nu }^{a}$.
One could of course also calculate everything in terms of the physical fields, however, we
emphasize again that in this case we need to consider the photon also in region 1). The
omission of the photon would lead to the violation of gauge invariance since the photon
contains a mixture of the $B_{\nu }$ and $W_{\nu }^{3}$ fields.

In region 1), the renormalization group equation (or generalized infrared evolution equation)
(\ref{eq:mrg}) in the case of all $m_i < M$ reads\footnote{We exclude here
top-Yukawa couplings which couple proportional to $\frac{m_t^2}{M^2}$ since
they don't have an analogue in QCD. It is, however, not unlikely that those
terms can also be included in the splitting functions fulfilling
Altarelli-Parisi equations. Note also, that the amplitude on the right hand
side is in general a linear combination of fields in the unbroken phase according to Eqs. (\ref{eq:wpm}),
(\ref{eq:z}) and (\ref{eq:p}). In addition, in the electroweak theory matching will be required at the
scale $M$ and often on-shell renormalization of the couplings $e$ and $\sin \theta_{\rm w}$ is used. In 
this case one has additional complications in the running coupling terms due to the different mass
scales involved below $M$. Details are presented in section \ref{sec:er}.}
\begin{eqnarray}
&& \left( \frac{\partial}{\partial t} + \beta \frac{\partial}{\partial g} + 
\beta^\prime \frac{\partial}{\partial g^\prime} + \sum_{i=1}^{n_g}
\Gamma^i_g(t) -n_{\rm W} \frac{1}{2} \frac{\alpha}{\pi} \beta_0
-n_{\rm B} \frac{1}{2} \frac{\alpha^\prime}{\pi} \beta^\prime_0 + \sum_{k=1}^{n_f} \left(
\Gamma^k_f(t) + \frac{1}{2} 
\gamma^k_{q\overline{q}} \right) \right) \nonumber \\
&& \times {\cal M}^\perp (p_1,...,p_n,g,g^\prime,\mu) = 0 
\label{eq:wmrg}
\end{eqnarray}
where the index $\perp$ indicates that we consider only $n_g$ transversely 
polarized external gauge bosons and $n_{\rm W}+n_{\rm B}=n_g$.
The two $\beta$-functions are given by:
\begin{eqnarray}
\beta (g(\overline{\mu}^2)) &=& \frac{\partial g(\overline{\mu}^2)}{\partial \log \overline{\mu}^2}
\approx - \beta_0 \frac{g^3(\overline{\mu}^2)}{8\pi^2} \label{eq:b0} \\
\beta^\prime (g^\prime(\overline{\mu}^2)) &=& \frac{\partial g^\prime
(\overline{\mu}^2)}{\partial \log \overline{\mu}^2}
\approx - \beta^\prime_0 \frac{{g^\prime}^3(\overline{\mu}^2)}{8\pi^2} \label{eq:b0p} 
\end{eqnarray}
with the one-loop terms given by:
\begin{equation}
\beta_0=\frac{11}{12}C_A - \frac{1}{3}n_{gen}-\frac{1}{24}n_{h} \;\;\;,\;\;\;
\beta^\prime_0= - \frac{5}{9}n_{gen} -\frac{1}{24}n_{h}
\end{equation}
where $n_{gen}$ denotes the number of fermion generations \cite{wein,gross} and $n_h$ the number of
Higgs doublets.
The infrared singular anomalous dimensions read
\begin{equation}
\Gamma^i_{f,g} (t) = \left( \frac{\alpha}{4 \pi}T_i(T_i+1)+ \frac{\alpha^\prime}{4 \pi} \left( \frac{Y_i}{2} 
\right)^2 \right) t
\label{eq:wad}
\end{equation}
where $T_i$ and $Y_i$ are the total weak isospin and hypercharge respectively of the particle emitting
the soft and collinear gauge boson. Analogously,
\begin{equation}
\gamma^i_{q\overline{q}} = -3 \left( \frac{\alpha}{4 \pi}T_i(T_i+1)+ \frac{\alpha^\prime}{4 \pi} \left( \frac{Y_i}{2}
\right)^2 \right) 
\label{eq:qad}
\end{equation}
The initial condition for Eq. (\ref{eq:wmrg}) is given by the requirement that for the infrared cutoff
$\mu^2=s$ we obtain the Born amplitude. The solution of (\ref{eq:wmrg}) is thus given by
\begin{eqnarray}
&& {\cal M}^\perp (p_1,...,p_n,g,g^\prime,\mu) = {\cal M}^\perp_{\rm Born} 
(p_1,...,p_n,g(s), g^\prime(s)) 
\nonumber \\ && \times \exp \left\{ - \frac{1}{2} \sum^{n_g}_{i=1} 
\left( \frac{\alpha(s)}{4 \pi}T_i(T_i+1)+ \frac{\alpha^\prime(s)}{4 \pi} 
\left( \frac{Y_i}{2} \right)^2 \right) \log^2 \frac{s}{\mu^2} \right. \nonumber \\
&& \;\;\;\;\;\;\;\;\;\;\;\;
+ \left( n_{\rm W} \frac{\alpha(s)}{2\pi} \beta_0 + n_{\rm B}
\frac{\alpha^\prime(s)}{2\pi} \beta^\prime_0 \right) \log \frac{s}{\mu^2} \nonumber \\
&& -\frac{1}{2} \left. \sum^{n_f}_{k=1} \left( \frac{\alpha(s)}{4 \pi}T_k(T_k+1)+ 
\frac{\alpha^\prime(s)}{4 \pi} \left( \frac{Y_k}{2} \right)^2 \right)
\left[ \log^2 \frac{s}{\mu^2} - 3 \log \frac{s}{\mu^2} \right] \right\} \label{eq:mpsol1}
\end{eqnarray}
where $n_{\rm W}$ and $n_{\rm B}$ denote the number of external $W$ and $B$ fields respectively.
The $SU(2) \times U(1)$ group factors in the exponential can be written in terms of     
the parameters of the broken theory as follows:     
\[     
g^{2}T_i(T_i+1)+{g^{\prime }}^{2}\left( \frac{Y_i}{2}\right)     
^{2}=e_i^{2}+g^{2}\left( T_i(T_i+1)-(T_i^{3})^{2}\right) +\frac{g^{2}}{\cos     
^{2}\theta _{w}}\left( T_i^{3}-\sin ^{2}\theta _{w}Q_i\right) ^{2},     
\]     
where the three terms on the r.h.s. correspond to the contributions of the     
soft photon (interacting with the electric charge $e_i=Q_ig\sin \theta _{w}$),     
the $W^{\pm }$ and the $Z$ bosons, respectively. Although we may rewrite     
solution (\ref{eq:mpsol1}) in terms of the parameters of the broken theory in     
the form of a product of three exponents corresponding to the exchanges of     
photons, $W^{\pm }$ and $Z$ bosons, it would be wrong to identify the     
contributions of the diagrams without virtual photons with this expression     
for the particular case $e_i^{2}=0$. This becomes evident when we note that if     
we were to omit photon lines then the result would depend on the choice of     
gauge, and therefore be unphysical. Only for $\theta _{w}=0$, where the     
photon coincides with the $B$ gauge boson, would the identification of the     
$%
e_i^{2}$ term with the contribution of the diagrams with photons be correct.     

We now need to discuss the solution in the general case. In region 1) we calculated the
scattering amplitude for the theory in the unbroken phase in the massless limit.
Choosing the cutoff $\mu$ in region 2), $\mu \ll M$, we have to only consider the
photon contribution. In this region we cannot necessarily neglect all mass terms, so
we need to discuss the subleading terms for QED with mass effects.
If $m_i\ll \mu$, the results from the massless QCD discussion of section \ref{sec:ap} can
be used directly by using the Abelian limit $C_F =1$. In case $\mu \ll m_i$ we must use the
well known next to leading order QED results, e.g. \cite{yfs}, and the virtual probabilities take the
following form for fermions:
\begin{equation}
w^f_i(s,\mu^2) = \left\{ \begin{array}{lc} \frac{e_i^2}{(4 \pi)^2} \left( \log^2 \frac{s}{\mu^2}
- 3 \log \frac{s}{\mu^2} \right) & , \;\;\; m_i \ll \mu \\
\frac{e_i^2}{(4 \pi)^2} \left[ \left( \log \frac{s}{m_i^2}-1 \right) 2 \log \frac{m_i^2}{\mu^2} \right. \\
\left.\;\;\;\;\;\;\;\;\;\;+ \log^2 \frac{s}{m_i^2} - 3 \log \frac{s}{m_i^2} \right] & , \;\;\; \mu \ll m_i\end{array} \right.
\end{equation}
Note, that in the last equation the full subleading collinear logarithmic term \cite{m} 
is used in distinction
to Ref. \cite{yfs}. In the explicit two loop calculation presented in Ref.
\cite{bnb} it can be seen that the full collinear term also exponentiates
at the subleading level in massive QED.
For $W^\pm$ bosons we have analogously:
\begin{equation}
w^{\rm w}_i(s,\mu^2) = 
\frac{e_i^2}{(4 \pi)^2} \left[ \left( \log \frac{s}{M^2}-1 \right)
2 \log \frac{M^2}{\mu^2}
+ \log^2 \frac{s}{M^2} \right]  
\end{equation}
In addition we have collinear terms for external on-shell photon lines
\footnote{I thank the authors of Ref. \cite{dp} for clarifying this point.}
from fermions with mass $m_j$ and electromagnetic charge $e_j$ 
up to scale $M$:
\begin{equation}
w_i^\gamma(M^2,\mu^2) = \left\{ \begin{array}{lc} 
\frac{n_f}{3} \frac{e_j^2}{4 \pi^2} N^j_C 
\log \frac{M^2}{\mu^2} & , \;\;\; m_j \ll \mu \\
\frac{1}{3} \sum_{j=1}^{n_f} \frac{e_j^2}{4 \pi^2} N^j_C \log \frac{M^2}{m_j^2}
& , \;\;\; \mu \ll m_j\end{array} \right.
\end{equation}
Note that automatically, $w_i^\gamma(M^2,M^2)=0$. At one loop order,
this contribution cancels against terms from the renormalization of the QED
coupling up to scale $M$. For external $Z$-bosons, however, there are no
such collinear terms since the mass is large compared to the $m_i$. Thus,
the corresponding RG-logarithms up to scale $M$ remain uncanceled.

The appropriate initial condition is given by Eq. (\ref{eq:mpsol1})
evaluated at the matching point $\mu=M$. Thus we find for the general solution in region 2):
\begin{eqnarray}
&& {\cal M}^\perp (p_1,...,p_n,g,g^\prime,\mu) = {\cal M}^\perp_{\rm Born} 
(p_1,...,p_n,g(s),g^\prime(s)) 
\nonumber \\ && \times \exp \left\{ - \frac{1}{2} \sum^{n_g}_{i=1} 
\left( \frac{\alpha(s)}{4 \pi}T_i(T_i+1)+ \frac{\alpha^\prime(s)}{4 \pi} 
\left( \frac{Y_i}{2} \right)^2 \right) \log^2 \frac{s}{M^2} \right. \nonumber \\
&& \;\;\;\;\;\;\;\;\;\;\;\;
+ \left( n_{\rm W} \frac{\alpha(s)}{2\pi} \beta_0 + n_{\rm B}
\frac{\alpha^\prime(s)}{2\pi} \beta^\prime_0 \right) \log \frac{s}{M^2} \nonumber \\
&& -\frac{1}{2} \left. \sum^{n_f}_{k=1} \left( \frac{\alpha(s)}{4 \pi}T_k(T_k+1)+ 
\frac{\alpha^\prime(s)}{4 \pi} \left( \frac{Y_k}{2} \right)^2 \right)
\left[ \log^2 \frac{s}{M^2} - 3 \log \frac{s}{M^2} \right] \right\} \nonumber \\
&&\times \exp \left[ - \frac{1}{2} \sum_{i=1}^{n_f} \left( w^f_i(s,\mu^2) 
- w^f_i(s,M^2) \right)  
- \frac{1}{2} \sum_{i=1}^{n_w} \left( w^{\rm w}_i(s,\mu^2) 
- w^{\rm w}_i(s,M^2) \right) \right. \nonumber \\
&& \;\;\;\;\;\;\;\;\;\;\; \left. -\frac{1}{2} \sum_{i=1}^{n_\gamma} w_i^\gamma(M^2,\mu^2) 
\right] \nonumber \\ 
&& = {\cal M}^\perp_{\rm Born} 
(p_1,...,p_n,g(s),g^\prime(s)) 
\nonumber \\ && \times \exp \left\{ - \frac{1}{2} \sum^{n_g}_{i=1} 
\left( \frac{\alpha(s)}{4 \pi}T_i(T_i+1)+ \frac{\alpha^\prime(s)}{4 \pi} 
\left( \frac{Y_i}{2} \right)^2 \right) \log^2 \frac{s}{M^2} \right. \nonumber \\
&& \;\;\;\;\;\;\;\;\;\;\;\;
+ \left( n_{\rm W} \frac{\alpha(s)}{2\pi} \beta_0 + n_{\rm B}
\frac{\alpha^\prime(s)}{2\pi} \beta^\prime_0 \right) \log \frac{s}{M^2} \nonumber \\
&& - \frac{1}{2} \left. \sum^{n_f}_{k=1} \left( \frac{\alpha(s)}{4 \pi}T_k(T_k+1)+ 
\frac{\alpha^\prime(s)}{4 \pi} \left( \frac{Y_k}{2} \right)^2 \right)
\left[ \log^2 \frac{s}{M^2} - 3 \log \frac{s}{M^2} \right] \right\} \nonumber \\
&&\times \exp \left[ - \frac{1}{2} \sum_{i=1}^{n} \! \left( \frac{e_{i}^{2}(s)}{(4\pi )^{2}}%
\! \left( 2 \log \frac{s}{m_{i}\,M}\log \frac{M^{2}}{m_{i}^{2}}+ 2 \log \frac{s}{%
m_{i}^{2}}\log \frac{m_{i}^{2}}{\mu ^{2}} + 3 \log 
\frac{m_i^2}{M^2} \right.  \right.  \right.  \nonumber \\
&& \left. \left. \left. -2 \log \frac{m_i^2}{\mu^2} 
\right) \right) - \frac{1}{2} \sum_{i=1}^{n_\gamma} \frac{1}{3} 
\frac{e_i^2}{4 \pi^2} N^i_C \log \frac{M^2}{m_i^2} \right] \label{eq:mpsol2}
\end{eqnarray}
The last equality holds for $\mu \ll m_i \ll M$ and we have absorbed all $\beta$-function terms
{\it not} related to external lines
into redefinitions of the scales of the couplings. 
It is important to note again that, unlike the situation in QCD, in the electroweak theory we
have in general different mass scales determining the running of the couplings of the physical on-shell
renormalization scheme quantities. We have written the above result in such
a way that it holds for arbitrary chiral fermions and transversely polarized gauge bosons. 
In order to include physical external photon states in the on-shell scheme, 
the renormalization condition is given by the requirement that the physical
photon does not mix with the Z-boson. This leads to the condition that
the Weinberg rotations in Fig. \ref{fig:mix} at one loop receive no
RG-corrections. Thus, above the scale $M$ the subleading collinear and
RG-corrections cancel for physical photon and Z-boson states.
For physical observables, soft real photon emission must be taken into account in an inclusive (or
semi inclusive) way and the parameter $\mu^2$ in (\ref{eq:mpsol2}) will be replaced by parameters
depending on the experimental requirements. This will be briefly discussed in the following section.

\subsubsection{Semi inclusive cross sections} \label{sec:si}

In order to make predictions for observable cross sections, the unphysical infrared cutoff $\mu^2$
has to be replaced with a cutoff $\mu^2_{exp}$, related to the lower bound of
$\mbox{\boldmath $k$}_\perp^2$ of the other virtual particles     
of those gauge bosons emitted in the process which are not included in the cross section.
We assume that $\mu^2_{exp}< M^2$, so that the non-Abelian component of the photon is not essential.
The case $\mu^2_{exp}>M^2$ is much more complicated and is discussed in Ref. \cite{flmm} through
two loops at the DL level.

We again only discuss transversely polarized external gauge boson in the Born process and can write
the expression for the semi-inclusive cross section:
\begin{eqnarray}     
d\sigma^\perp (p_{1}, \ldots, p_{n},g,g^\prime,\mu_{exp}) &=& d\sigma^\perp_{\rm elastic}(p_{1}, 
\ldots ,p_{n},g(s),g^\prime (s),\mu) \nonumber \\ && \times \exp   
(w_{\exp}^{\gamma }(s,m_i,\mu,\mu_{exp}))     
\end{eqnarray}     
In the soft photon approximation we have:
\begin{eqnarray}
w_{\exp}^{\gamma }(s,m_i,\mu,\mu_{exp}) \!\!&=&\!\!\! \left\{ \begin{array}{lc} 
\frac{e_i^2}{(4 \pi)^2} \sum_{i=1}^n \left[ 
- \log^2 \frac{s}{\mu^2_{exp}} 
+ \log^2 \frac{s}{\mu^2}- 3 \log \frac{s}{\mu^2} \right] 
& , m_i \ll \mu \\
\frac{e_i^2}{(4 \pi)^2} \sum_{i=1}^n \left[ \left( \log 
\frac{s}{m_i^2} -  1 \right) 
2 \log \frac{m_i^2}{\mu^2} + \log^2 \frac{s}{m_i^2} 
\right. \\ \left. - 2 \log \frac{s}{\mu^2_{exp}} \left( \log \frac{s}{m_i^2} -
1 \right) \right]
& , 
\mu \ll m_i \end{array} \right. \nonumber \\ &&
\end{eqnarray}
where the upper case applies only to fermions since for $W^\pm$ 
we have $\mu < M$ in region 2).
Since the upper bound on ${\mbox{\boldmath $k$}_{\perp }^{2}}$ of the     
photons which are allowed to be radiated is less than $M^{2}$, we must use the 
cut-off $\mu ^{2}<M^{2}$ and, consequently, (\ref{eq:mpsol2}) for the matrix element of  
the non-radiative process.  Therefore, we obtain\footnote{The notation here is again simplified in the
sense that for $Z$-boson and $\gamma$ final states one has to include the mixing correctly as described
above.}
\begin{eqnarray}     
&& d\sigma^\perp (p_{1}, \ldots, p_{n},g,g^\prime,\mu_{exp}) = d\sigma_{\rm Born}^\perp(p_{1},
\ldots ,p_{n},g(s),g^\prime (s))
\nonumber \\ && \times \exp \left\{ - \sum^{n_g}_{i=1} 
\left( \frac{\alpha(s)}{4 \pi}T_i(T_i+1)+ \frac{\alpha^\prime(s)}{4 \pi} 
\left( \frac{Y_i}{2} \right)^2 \right) \log^2 \frac{s}{M^2} \right. \nonumber \\
&& \;\;\;\;\;\;\;\;\;\;\;\;
+ \left( n_{\rm W} \frac{\alpha(s)}{\pi} \beta_0 + n_{\rm B}
\frac{\alpha^\prime(s)}{\pi} \beta^\prime_0 \right) \log \frac{s}{M^2} \nonumber \\
&& \left. - \sum^{n_f}_{k=1} \left( \frac{\alpha(s)}{4 \pi}T_k(T_k+1)+ 
\frac{\alpha^\prime(s)}{4 \pi} \left( \frac{Y_k}{2} \right)^2 \right)
\left[ \log^2 \frac{s}{M^2} - 3 \log \frac{s}{M^2} \right] \right\} \nonumber \\
&&\times \exp \left[ - \sum_{i=1}^{n_f} \left( w^f_i(s,\mu^2) 
- w^f_i(s,M^2) \right)  
- \sum_{i=1}^{n_w} \left( w^{\rm w}_i(s,\mu^2) 
- w^{\rm w}_i(s,M^2) \right) \right. \nonumber \\ 
&& \;\;\;\;\;\;\;\;\;\;\; \left. - \sum_{i=1}^{n_\gamma} w_i^\gamma(M^2,m_j^2) 
\right] \nonumber \\ 
&& \times \exp \left(
\frac{e_i^2}{(4 \pi)^2} \sum_{i=1}^n \left[ \left( \log 
\frac{s}{m_i^2} - 1 \right) 2 \log \frac{m_i^2}{\mu^2} - 2 
\log \frac{s}{\mu^2_{exp}}\left( \log \frac{s}{m_i^2} - 1 \right)
\right. \right. \nonumber \\ && \left. \left.
+ \log^2 \frac{s}{m_i^2} \right] \right) \label{eq:si}
\end{eqnarray}     
where we use $\mu \ll m_i$.
The $\mu$ dependence in this expression cancels and the semi-inclusive cross section depends only on
the parameters of the experimental requirements.

Eq. (\ref{eq:si}) contains all leading double and single logarithms to cross sections\footnote{We emphasize
again that we did not consider angular logarithms which can be sizable and should be calculated
at least to one loop order.} containing
arbitrary numbers of external fermions and transversely polarized gauge bosons. We have only assumed
that all masses are not larger than the electroweak scale $M$ and impose a cut on the the allowed
values of emitted real gauge bosons 
${\mbox{\boldmath $k$}_{\perp }^{2}} \leq \mu^2_{exp} < M^2$, i.e. up to the weak scale
we only need to consider real QED effects.

\subsection{Longitudinal degrees of freedom} \label{sec:ld}

In this section we discuss if results obtained from the massless unbroken phase of the $SU(2)\times U(1)$
theory,
where due to gauge invariance we have only transverse physical degrees of freedom, 
can be extended to the full theory including longitudinal vector bosons. This point of discussion is
necessary and important since the longitudinal degrees of freedom don't decouple at high energies and
could give crucial clues to potentially strong dynamical effects for large Higgs masses $m_H \sim 1 TeV$
\cite{wein}.

The connection between the strategy pursued for the transverse degrees of freedom and the 
corrections to longitudinally polarized vector bosons at high energies is provided by the Goldstone
boson equivalence theorem \cite{gb}. It states that at tree level for S-matrix elements for longitudinal
bosons at the high
energy limit $M^2/s\longrightarrow 0$ can be expressed through matrix elements involving their associated
would be Goldstone bosons. We write schematically in case of a single gauge boson:
\begin{eqnarray}
{\cal M}(W^\pm_{L}, \psi_{{\rm phys}}) &=& {\cal M}(\phi^\pm, \psi_{{\rm phys}}) + {\cal O} \left( 
\frac{M_{\rm w}}{\sqrt{s}} \right)
\label{eq:wet} \\
{\cal M}(Z_{L}, \psi_{{\rm phys}}) &=& {\cal M}(\phi, \psi_{{\rm phys}}) + {\cal O} \left(
\frac{M_{\rm z}}{\sqrt{s}} \right)
\label{eq:zet}
\end{eqnarray}
The problem with this statement of the equivalence theorem is that it holds only at tree level
\cite{yy,bs}. For
calculations at higher orders, 
additional terms enter which change Eqs. (\ref{eq:wet}) and (\ref{eq:zet}).

Because of the gauge invariance of the physical theory and the associated BRST invariance, a modified
version of Eqs. (\ref{eq:wet}) and (\ref{eq:zet}) can be derived \cite{yy} which reads
\begin{eqnarray}
k^\nu {\cal M}(W^\pm_{\nu}(k), \psi_{{\rm phys}}) &=& C_{\rm w} M_{\rm w} {\cal M}(\phi^\pm (k), 
\psi_{{\rm phys}}) + {\cal O} \left( \frac{M_{\rm w}}{\sqrt{s}} \right) \label{eq:wetp} \\
k^\nu{\cal M}(Z_{\nu}(k), \psi_{{\rm phys}}) &=& C_{\rm z} M_{\rm z} {\cal M}(\phi (k), \psi_{{\rm phys}})
+ {\cal O} \left( \frac{M_{\rm z}}{\sqrt{s}} \right) \label{eq:zetp}
\end{eqnarray}
where the multiplicative factors $C_{\rm w}$ and $C_{\rm z}$ depend only on wave function renormalization
constants and mass counterterms. Thus, using the form of the longitudinal polarization vector of 
Eq. (\ref{eq:lpv}) we can write
\begin{eqnarray}
{\cal M}(W^\pm_{L}(k), \psi_{{\rm phys}}) &=&  C_{\rm w} {\cal M}(\phi^\pm (k), 
\psi_{{\rm phys}}) + {\cal O} \left( \frac{M_{\rm w}}{\sqrt{s}} \right) \label{eq:wgs} \\
{\cal M}(Z_{L}(k), \psi_{{\rm phys}}) &=&  C_{\rm z} {\cal M}(\phi (k), \psi_{{\rm phys}})
+ {\cal O} \left( \frac{M_{\rm z}}{\sqrt{s}} \right) \label{eq:zgs}
\end{eqnarray}
Thus we see that in principle, 
there are logarithmic loop corrections to the tree level equivalence theorem\footnote{
An exception is the background field gauge where the Ward-identities guarantee that the factors
$C_{\rm w}=1$ and $C_{\rm z}=1$ to all orders \cite{dd}. It should thus
be investigated if subleading
corrections can also be obtained from the Goldstone boson 
equivalence theorem.}. In addition, for longitudinal gauge bosons we also have logarithmic
corrections with Yukawa terms \cite{bddms}.
On the one hand, this means that the method of section
\ref{sec:td} should be used with caution to obtain the
relevant subleading terms. Thus we should consider these
corrections separately. On the other, since
the corrections are at most logarithmic, 
it means that the results of Ref. \cite{flmm} can be extended to
the longitudinal sector as well.
Thus we 
find\footnote{For longitudinally polarized 
$Z$-boson final states there are no mixing terms since the photon has only transverse polarization 
states. Thus one needs to only include the associated Goldstone boson $\phi$ at the DL level.}
for $\mu_{exp} < M$:
\begin{eqnarray}     
&& d\sigma^\parallel (p_{1}, \ldots, p_{n},g,g^\prime,\mu_{exp}) = d\sigma_{\rm Born}^\phi(p_{1},
\ldots ,p_{n},g,g^\prime)
\nonumber \\ && \times \exp \left\{ - \sum^{n}_{i=1} 
\left( \frac{\alpha}{4 \pi}T_i(T_i+1)+ \frac{\alpha^\prime}{4 \pi} 
\left( \frac{Y_i}{2} \right)^2 \right) \log^2 \frac{s}{M^2} \right\} \nonumber \\
&&\times \exp \left[ - \sum_{i=1}^{n} \left( w^{{\rm DL}}_i(s,\mu^2) 
- w^{{\rm DL}}_i(s,M^2) \right)  
+ w^{{\rm DL}}_{exp}\right] \label{eq:sipa}
\end{eqnarray}     
where the index $\parallel$ indicates the cross section for longitudinally polarized gauge bosons, 
while the field $\phi$ indicates that the appropriate fields and
quantum numbers on the r.h.s. in Eq. (\ref{eq:sipa})
are those of the associated would be Goldstone bosons.

Thus, we have shown that all DL corrections can be summed to all orders by employing the
evolution equation approach of Ref. \cite{flmm} in connection with the Goldstone boson equivalence
theorem.

\section{Comparison with explicit results} \label{sec:er}

In this section we compare our results obtained in the previous sections with known results in 
special cases and one loop calculations. In Ref. \cite{kps}, QCD-results for the Sudakov form
factor were generalized to the high energy electroweak theory\footnote{In addition, angular terms
at the one loop level were calculated which we do not consider in this work.}. Since the general
strategy pursued is the same as in Ref. \cite{flmm}, we of course agree with their result for
leading and subleading electroweak corrections to $e^+ e^- \longrightarrow f \overline{f}$ to all orders.

\begin{figure} 
\centering 
\epsfig{file=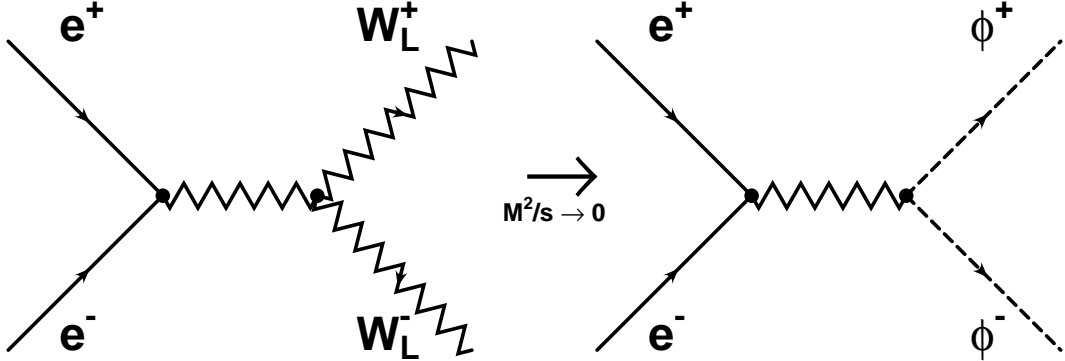,width=14cm} 
\caption{The pictorial Goldstone boson equivalence theorem for $W$-pair production in $e^+e^-$ collisions.
The correct DL-asymptotics for longitudinally polarized bosons 
are obtained by using the quantum numbers of the charged would be Goldstone scalars at high energies.}
\label{fig:gsb}
\end{figure} 
A very important check is provided by the explicit one-loop corrections of Ref. \cite{bddms} for 
high energy on-shell $W$-pair production in the soft photon approximation. 
In the following, the lower index on the cross section indicates the helicity of the electron, where
$e^-_-$ denotes the left handed electron.
We summarize the relevant
results for $e^+_+ e^-_- \longrightarrow W^+_\perp W^-_\perp$,
$e^+_+ e^-_- \longrightarrow W^+_\parallel W^-_\parallel$ and $e^+_- e^-_+ \longrightarrow W^+_\parallel
W^-_\parallel$ for convenience as follows:
\begin{eqnarray}
\left( \frac{ d \sigma}{d \Omega} \right)_{-,\, \perp} \!\! &\approx& \!\! \left( \frac{d \sigma}{d \Omega} 
\right)^{\rm \!\! Born}_{-,\, \perp} \!\! \left\{ 1 + \frac{e^2}{8\pi^2} \left[ - \frac{1+2c_{\rm w}^2+8c_{\rm w}^4}{
4 c_{\rm w}^2 s_{\rm w}^2} \log^2 \frac{s}{M^2} 
+ 3 \frac{1-2c_{\rm w}^2+4c_{\rm w}^4}{4 c_{\rm w}^2 s_{\rm w}^2} \log \frac{s}{M^2}  
\right. \right. \nonumber \\ && \left. \left.
+ 3 \log \frac{s}{m_e^2} + 2 \log \frac{4 \Delta E^2}{s} \left( \log \frac{s}{m_e^2} + \log \frac{s}{
M^2}-2 \right) \right. \right. \nonumber \\
&& \left. \left. - \frac{4}{3} \sum_{j=1}^{n_f} Q_j^2 N^j_C \log \frac{m_j^2}{M^2} \right] \right\} \label{eq:mt} \\
\left( \frac{ d \sigma}{d \Omega} \right)_{-,\, \parallel} &\approx& \left( \frac{d \sigma}{d \Omega} 
\right)^{\rm Born}_{-,\, \parallel} \left\{ 1 + \frac{e^2}{8\pi^2} \left[ - \frac{1-2c_{\rm w}^2+4c_{\rm w}^4}{
2 c_{\rm w}^2 s_{\rm w}^2} \log^2 \frac{s}{M^2} 
\right. \right. \nonumber \\ && \left. \left.
+ 2 \log \frac{4 \Delta E^2}{s} \left( \log \frac{s}{m_e^2} + \log \frac{s}{
M^2} \right) \right] \right\} \label{eq:ml} \\
\left( \frac{ d \sigma}{d \Omega} \right)_{+,\, \parallel} &\approx& \left( \frac{d \sigma}{d \Omega} 
\right)^{\rm Born}_{+,\, \parallel} \left\{ 1 + \frac{e^2}{8\pi^2} \left[ - \frac{5-10c_{\rm w}^2+8c_{\rm w}^4}{
4 c_{\rm w}^2 s_{\rm w}^2} \log^2 \frac{s}{M^2} 
\right. \right. \nonumber \\ && \left. \left.
+ 2 \log \frac{4 \Delta E^2}{s} \left( \log \frac{s}{m_e^2} + \log \frac{s}{
M^2} \right) \right] \right\} \label{eq:pl}
\end{eqnarray}
where the last line in Eq. (\ref{eq:mt}) corresponds to a sum over all fermions contributing to the
coupling renormalization (with multiplicity $N_C=3$ for quarks and $N_C=1$ for leptons). 
This contribution can be included in the scale of the running on-shell charge $\alpha_{\rm eff}(M^2)$
\cite{kd}.
For the longitudinal cross sections we are only concerned with DL corrections.
The Born cross sections are given by:
\begin{eqnarray}
\left( \frac{d \sigma}{d \Omega} 
\right)^{\rm Born}_{-,\, \perp} &=& \frac{e^4}{64 \pi^2s} \frac{1}{4 s_{\rm w}^4} \frac{u^2+t^2}{t^2}
\sin^2 \theta \label{eq:bmt} \\
\left( \frac{d \sigma}{d \Omega} \right)^{\rm Born}_{-,\, \parallel} &=& \frac{e^4}{64 \pi^2s} 
\frac{1}{16 s_{\rm w}^4 c_{\rm w}^4} \sin^2 \theta \label{eq:bmp} \\
\left( \frac{d \sigma}{d \Omega} \right)^{\rm Born}_{+,\, \parallel} &=& \frac{e^4}{64 \pi^2s} 
\frac{1}{4 c_{\rm w}^4} \sin^2 \theta \label{eq:bpp} 
\end{eqnarray}
where we keep the angular dependence. 
In Eq. (\ref{eq:bmt}) a sum over the two transverse polarizations of the $W^\pm$ ($++$ and $--$) 
is implicit.
These expressions demonstrate that the longitudinal cross sections
in Eqs. (\ref{eq:bmp}) and (\ref{eq:bpp}) are not suppressed with respect to Eq. (\ref{eq:bmt}). On the other
hand, $\left( \frac{d \sigma}{d \Omega} \right)^{\rm Born}_{+,\, \perp}$ is mass suppressed \cite{bddms}.

Eqs. (\ref{eq:mt}), (\ref{eq:ml}) and (\ref{eq:pl}) were of course calculated in terms of the physical fields
of the broken theory and in the on-shell scheme. We denote $c_{\rm w}=\cos \theta_{
\rm w}$ and $s_{\rm w}=\sin \theta_{\rm w}$ respectively.
In order to compare with the results of section \ref{sec:bg} we listed the relevant quantum numbers in
Table \ref{tab:qn}. For comparison with Eq. (\ref{eq:mt}) and to logarithmic accuracy, 
we can absorb the running coupling effects from our massless
scheme to the on-shell scheme as follows. The Born cross section
in our approach is proportional to $g^4$ (see Eq. (\ref{eq:bmt})). The coupling renormalization
above the scale $M$ is given by:
\begin{equation}
g^2(s)=g^2(M^2) \left( 1- \frac{g^2(M^2)}{4\pi^2} \left( \frac{11}{12} C_A
- \frac{1}{24} n_h - \frac{n_{gen}}{3} 
\right) \log \frac{s}{M^2} \right) \label{eq:rc}
\end{equation}
Below the scale where non-Abelian effects enter, the running is only due to the electromagnetic coupling
and we write $g^2(M^2)=\frac{e^2_{\rm eff}(M^2)}{s_{\rm w}^2}$ with
\begin{equation}
e^2_{\rm eff}(M^2)=e^2 \left(1+ \frac{1}{3} \frac{e^2}{4 \pi^2} \sum_{j=1}^{n_f} Q_j^2 N^j_C \log \frac{M^2}{m_j^2} \right)
\end{equation}
We therefore observe that the running coupling terms proportional to $\log \frac{s}{M^2}$ 
cancel for this process with the subleading contributions
from the virtual splitting functions (see Eq. (\ref{eq:mpsol2})) and what remains are just the
Abelian terms up to scale $M$.
Thus for Eq. (\ref{eq:mt}) we obtain from Eq. (\ref{eq:si}) at the one loop level:
\begin{eqnarray}
\left( \frac{ d \sigma}{d \Omega} \right)_{-,\, \perp}\!\! &=& \!\! \left( \frac{d \sigma}{d \Omega} 
\right)^{\rm Born}_{-,\, \perp} \left\{ 1 - \left( \frac{g^2}{8 \pi^2} T_{\rm w} (T_{\rm w}+1)
+ \frac{{g^\prime}^2}{8 \pi^2} \frac{Y^2_{\rm w}}{4} \right) \log^2 \frac{s}{M^2} \right. \nonumber \\
\!\! && \!\! - \left( \frac{g^2}{8 \pi^2} T_{e^-_-} (T_{e^-_-}+1) + \frac{{g^\prime}^2}{8 \pi^2} \frac{Y^2_{e^-_-}}{4}
\right) \left( \log^2 \frac{s}{M^2}- 3 \log \frac{s}{M^2} \right) - \frac{e^2}{8 \pi^2} \times \nonumber \\
\!\! && \!\! \left[ \left( \log \frac{s}{m_e^2} - 1 \right) 2 \log \frac{m_e^2}{\mu^2}
+\log^2 \frac{s}{m_e^2}-3 \log \frac{s}{m_e^2}- \log^2 \frac{s}{M^2}+3 \log \frac{s}{M^2}\right.
\nonumber \\ \!\! && \!\! + 2 \left( \log \frac{s}{M^2}-1 \right) \log \frac{M^2}{\mu^2} - \left(
\log \frac{s}{m_e^2} - 1 \right) \left( 2 \log \frac{m_e^2}{\mu^2}-2 \log \frac{s}{\mu^2_{exp}} \right)
- \nonumber \\ \!\! && 2 \left( \log \frac{s}{M^2}-1 \right) 
\left( \log \frac{M^2}{\mu^2} - \log \frac{s}{\mu^2_{exp}} \right) - \log^2 \frac{s}{m_e^2} -
\log^2 \frac{s}{M^2} \Bigg]  
\nonumber \\ 
\!\! && \left. + \frac{2}{3} \frac{e^2}{4 \pi^2} \sum_{j=1}^{n_f} Q_j^2 N^j_C \log \frac{M^2}{m_j^2} \right\}
\nonumber \\
\!\! &=& \!\! \left( \frac{d \sigma}{d \Omega}\right)^{\rm \!\! Born}_{-,\, \perp} \!
\left\{ 1 - \frac{e^2}{8 \pi^2} \left( \frac{1+10 c_{\rm w}^2}{4 s_{\rm w}^2 c_{\rm w}^2}
\log^2 \frac{s}{M^2} 
- 3 \frac{1+2 c_{\rm w}^2}{4 s_{\rm w}^2 c_{\rm w}^2}
\log \frac{s}{M^2} \right) + \frac{e^2}{8 \pi^2} \times \right. \nonumber \\
\!\! && \!\! \left[ 2 \log^2 \frac{s}{M^2} - 3 \log \frac{m_e^2}{M^2} - 4 \log
\frac{s}{\mu^2_{exp}} \left( \log \frac{s}{m_e M}-1 \right) \right] \nonumber \\
\!\! && \left. + \frac{2}{3} \frac{e^2}{4 \pi^2} \sum_{j=1}^{n_f} Q_j^2 N^j_C \log \frac{M^2}{m_j^2} \right\}
\label{eq:mymt} 
\end{eqnarray}
Eq. (\ref{eq:mymt}) agrees with Eq. (\ref{eq:mt}), which are both valid in the
soft photon approximation. Here and below we assume that 
$\Delta E < M$ and $\mu_{exp} < M$.
Analogously in the DL approximation,
it is straightforward to check the validity of our results for Eqs. (\ref{eq:ml}) and
(\ref{eq:pl}), emphasizing again that in this case we need to use the quantum numbers of the associated
Goldstone bosons, see Fig. \ref{fig:gsb} and Tab. \ref{tab:qn}.
\begin{table}
\begin{center}
\begin{Large}
\begin{tabular}{|l|c|c|r|}
\hline
& T & Y & Q \\
\hline
$e^-_-$ & 1/2 & -1 & -1 \\
\hline
$e^-_+$ & 0 & -2 & -1 \\
\hline
$e^+_+$ & 1/2 & 1 & 1 \\
\hline
$e^+_-$ & 0 & 2 & 1 \\
\hline
$W^\pm$ & 1 & 0 & $\pm$1 \\
\hline
$\phi^\pm$ & 1/2 & $\pm$1 & $\pm$1 \\
\hline
\end{tabular}
\end{Large}
\end{center}
\caption{The quantum numbers of various particles in the electroweak theory. 
The indices indicate the helicity of the electrons. We neglect all mass terms.
For longitudinally
polarized gauge bosons, the associated scalar Goldstone bosons describe the DL asymptotics.}
\label{tab:qn}
\end{table}

Thus we have verified that our results, calculated in terms of the unbroken massless fields,
give the correct leading and subleading logarithms in transversely polarized $W$-pair production
at the one loop level. For longitudinally polarized $W$-pairs the correct DL asymptotics is 
reproduced.

\section{Conclusions}

In this paper we considered the calculation of virtual next to leading electroweak corrections at energies
much larger than the electroweak scale when all particle masses can be neglected. We follow the
same approach as in Ref. \cite{flmm} which consists of using the fields of the unbroken theory to obtain
logarithmic corrections with the infrared evolution equation method in different regions of the infrared
cutoff. When particle masses can be neglected there is a one to one correspondence between the high
and low energy scaling behavior and the evolution equation can be formulated in terms of the
renormalization group with infrared singular anomalous dimension. The next to leading kernel can then
be obtained from the virtual contribution to the Altarelli-Parisi splitting functions. 

For external gauge boson emission one can use the above approach for transverse degrees of freedom.
For fermions and $W^\pm$ external states, the next to leading corrections exponentiate with respect to
the physical Born amplitudes. For $Z$ boson and $\gamma$ final states, one needs to include the effect
of mixing appropriately. For these final states we have exponentiation with respect to the amplitudes
containing the fields of the unbroken theory but {\it not} with respect to the physical Born amplitude.

For longitudinal degrees of freedom, one can use the Goldstone-boson equivalence theorem to obtain the
correct DL asymptotic behavior. These terms are found to exponentiate as well.
Loop corrections, however, lead to additional corrections including Yukawa terms at the subleading
level and should be considered separately.

We also compared our results with exact one loop calculations in the high energy approximation and
could reproduce the leading and subleading terms for transversely polarized
$W$-pair production in $e^+e^-$ collisions and the DL corrections for the longitudinal degrees of freedom.

Finally we note that there are of course terms which we have neglected in this analysis. As mentioned
above, there
are angular logarithms of the form $\log \frac{s}{M^2} \log \frac{t}{u}$, which in general
could be significant and
should be computed separately. We also omitted
all mass-logarithms of the form $\log \frac{s}{M^2} \log \frac{M_Z}{M_W}$ and
top-Yukawa terms. For the latter, it might be possible to include them consistently
into the virtual splitting functions.
For longitudinal degrees of freedom it would also be very helpful to 
calculate subleading corrections.

In conclusion, for future high energy experiments in the multi-TeV energy regime, the leading high
energy behavior of general scattering amplitudes can be an important ingredient to study the effect
of new physics expected in precisely this range.

\section*{Acknowledgements}  
   
We would like to thank A.~Denner and S.~Pozzorini for valuable discussions, especially concerning mixing
effects and corrections to the Goldstone boson equivalence theorem. We would also like to thank
W.~Beenakker, V.S.~Fadin and D.~Graudenz for consultations.


\begin{thebibliography}{99}   
\bibitem{flmm} V.S.~Fadin, L.N.~Lipatov, A.D.~Martin, M.~Melles; 
Phys.Rev. {\bf D61} (2000) 094002.

\bibitem{kl}  R. Kirschner, L. N. Lipatov; JETP {\bf 83} (1982) 488;   
Phys. Rev. {\bf D26} (1982) 1202.   

\bibitem{ccc} M.~Ciafaloni, P.~Ciafaloni, D.~Comelli; Phys.~Rev.~Lett. {\bf 84}:4810, 2000.

\bibitem{cc} P. Ciafaloni, D. Comelli; Phys.~Lett.~{\bf B476} (2000) 49.

\bibitem{dp} A.~Denner, S.~Pozzorini; PSI-PR-00-15.
   
\bibitem{bddms} W.~Beenakker, A.~Denner, S.~Dittmaier, R.~Mertig, T.~Sack; Nucl. Phys. {\bf B410}
(1993) 245.

\bibitem{NA}  J. J. Carazzone, E. C. Poggio, H. R. Quinn, Phys. Rev. {\bf   
%
D11} (1975) 2286; \newline   
J. M. Cornwall, G. Tiktopoulos, Phys. Rev. Lett. {\bf 35} (1975) 338;  \newline 
V. V. Belokurov, N. I. Usyukina, Phys. Lett. {\bf B94} (1980) 251; \newline
Theor. Math. Phys. {\bf 44} (1980) 657; {\bf 45} (1980) 957; \newline
J.C.~Collins; Phys.~Rev. {\bf D22} (1980) 1478; A.~Sen; Phys.~Rev. {\bf D24} (1981) 3281; \newline
G.P.~Korchemsky; Phys.~Lett. {\bf B217} (1989) 330; Phys.~Lett. {\bf B220} (1989) 629.

\bibitem{vg}  V. N. Gribov, Yad. Fiz. 5 (1967) 399 (Sov. J. Nucl. Phys. {\bf 5} 
(1967) 280); Sov. J. Nucl. Phys. {\bf   
%
12} (1971) 543; \newline   
L. N. Lipatov, Nucl. Phys. {\bf B307} (1988) 705; \newline   
V. Del Duca, Nucl. Phys. {\bf B345} (1990) 369.   

\bibitem{s}  V. V. Sudakov, Sov. Phys. JETP {\bf 3} (1956) 65.   

\bibitem{m} M.~Melles; A.~Phys.~Pol. {\bf B 28} (1997) 1159.

\bibitem{col} J.~Frenkel, J.C.~Taylor; Nucl.~Phys. {\bf 116} (1976) 64; 
J.~Frenkel, R.~Meuldermans, Phys.~Lett. {\bf B 65} (1976) 64; J.~Frenkel, Phys.~Lett. {\bf B 65} (1976)
383; K.J.~Kim, University of Mainz Preprint MZ-TH 76/6.

\bibitem{ap} G.~Altarelli, G.~Parisi; Nucl.~Phys. {\bf B126} (1977) 298.

\bibitem{aem} G.~Altarelli, R.K.~Ellis, G.~Martinelli; Nucl.~Phys. {\bf B157} (1979) 461.

\bibitem{a} G.~Altarelli; Phys.~Rept. {\bf 81} (1982) 1.

\bibitem{dot} A.I.~Davydychev, P.~Osland, O.V.~Tarasov; Phys.~Rev. {\bf D54} (1996) 4087.

\bibitem{gl} V.N.~Gribov, L.N.~Lipatov; Yad.~Fiz.{\bf 15} (1972) 1218, Sov.~J.~Nucl.~Phys. {\bf 15} (1972)
675; 
Yad.~Fiz. {\bf 15} (1972) 781, Sov.~J.~Nucl.~Phys. {\bf 15} (1972) 438.

\bibitem{pqz} E.C.~Poggio, H.R.~Quinn, J.B.~Zuber; Phys.~Rev.~{\bf D15} (1977) 1630.

\bibitem{p} E.C.~Poggio; Phys.~Rev.~{\bf D16} (1977) 2586.

\bibitem{kr} G.P.~Korchemsky, A.V.~Radyushkin; Phys.~Lett. {\bf B 171} (1986) 459.

\bibitem{ct}
J. M. Cornwall, G. Tiktopoulos;   
Phys. Rev. {\bf D13} (1976) 3370.   

\bibitem{ar} C.P.~Korthals-Altes, E.de~Rafael; Nucl.~Phys. {\bf B106} (1976) 237.

\bibitem{ku} T.~Kinoshita, A.~Ukawa; Phys.~Rev. {\bf D16} (1977) 332.

\bibitem{sterm} G.~Sterman, ``An introduction to quantum field theory'', Cambridge University Press 1993.

\bibitem{wein} S.~Weinberg, ``The quantum theory of fields (II)'', Cambridge University Press 1996.

\bibitem{gross} G.G.~Gross, ``Grand Unified Theories'', Oxford University Press 1984.

\bibitem{yfs} D.R.~Yennie, S.C.~Frautschi, H.~Suura; Ann.~Phys. (NY) {\bf 13} (1961) 379.

\bibitem{bnb} F.A.~Berends, W.L.van~Neerven, G.J.H.~Burgers; 
Nucl.~Phys.~{\bf B 297}:429, 1988, Erratum-ibid. {\bf B 304}:921, 1988. 

\bibitem{gb} J.M.~Cornwall, D.N.~Levin, G.~Tiktopoulos; Phys.~Rev. {\bf D10} (1974) 1145;
C.E.~Vayonakis, Lett.~Nuov.~Cim. {\bf 17} (1976) 383; M.S.~Chanowitz, M.K.~Gaillard, Nucl.~Phys.
{\bf B 261} (1985) 379.

\bibitem{yy} Y.P.~Yao, C.P.~Yuan; Phys.~Rev. {\bf 38} (1988) 2237.

\bibitem{bs} J.~Bagger, C.~Schmidt; 
Phys.~Rev.~{\bf D41} (1990) 264. 

\bibitem{dd} A.~Denner, S.~Dittmaier; Phys.~Rev. {\bf D54} (1996) 4499.

\bibitem{kps} J.H.~K\"uhn, A.A.~Penin, V.A.~Smirnov; hep-ph/9912503.

\bibitem{kd} E.~de~Rafael, J.L.~Rosner; Ann.~Phys. {\bf 82} (1973) 369;
C.P.~Korthals-Altes, E.~de~Rafael; 
Nucl.~Phys. {\bf B106} (1976) 237. 
\end{thebibliography}
\end{document}